\documentclass[bibyear]{aa}
\usepackage[varg]{txfonts}
\usepackage{graphicx}   
\usepackage{amsmath}    
\usepackage{mathrsfs}
\usepackage{amssymb}

\usepackage{sidecap}
\usepackage[breaklinks, colorlinks, citecolor=blue, urlcolor = blue]{hyperref}
\usepackage{pdflscape}
\usepackage{float}
\usepackage{here}
\usepackage[export]{adjustbox}
\usepackage{array}
\usepackage{natbib}
\bibpunct{(}{)}{;}{a}{}{,} 

\newcommand{\nustar}{\textit{NuSTAR }} 
\newcommand{\xmm}{\textit{XMM-Newton }} 
\newcommand{\suzaku}{\textit{Suzaku }}
\newcommand{\logxi}{$\log(\xi / \rm erg\, \rm s^{-1}\,\rm cm)$ }
\newcommand{\flux}{$\rm erg\,\rm cm^{-2}\,\rm s^{-1}$ }

\newcommand{\ka}{K$\alpha$ }
\newcommand{\kb}{K$\beta$ }

\newcommand{\Fexxvi}{Fe\,\textsc{xxvi} }

\begin{document} 

\nolinenumbers

   \title{\xmm- \nustar monitoring campaign of the Seyfert 1 galaxy IC~4329A}

   \author{A. Tortosa\thanks{\email{alessia.tortosa@inaf.it}} \inst{1,2}
          \and C. Ricci \inst{2,3,4} 
          \and E. Shablovinskaia \inst{2} 
          \and F. Tombesi \inst{5,1,6,7,8} 
          \and T. Kawamuro \inst{9}
          \and E. Kara \inst{10} 
          \and G. Mantovani \inst{11}
          \and M. Balokovic \inst{12}
          \and C-S. Chang \inst{13}
          \and K. Gendreau \inst{8}
          \and M. J. Koss \inst{14,15}
          \and T. Liu \inst{16}
          \and M. Loewenstein \inst{8,17}
          \and S. Paltani \inst{18}
          \and G. C. Privon \inst{19,20,21}
          \and B. Trakhtenbrot \inst{22}
          }
          
    
   \institute{INAF – Astronomical Observatory of Rome, Via Frascati 33, 00040 Monte Porzio Catone, Italy
   \and Instituto de Estudios Astrof\'isicos, Facultad de Ingenier\'ia y Ciencias, Universidad Diego Portales, Av. Ej\'ercito Libertador 441, Santiago, Chile.
   \and Kavli Institute for Astronomy and Astrophysics, Peking University, Beijing 100871, China
   \and George Mason University, Department of Physics \& Astronomy, MS 3F3, 4400 University Drive, Fairfax, VA 22030, USA
   \and Physics Department, Tor Vergata University of Rome, Via della Ricerca Scientifica 1, 00133 Rome, Italy
   \and INFN - Rome Tor Vergata, Via della Ricerca Scientifica 1, 00133 Rome, Italy 
   \and Department of Astronomy, University of Maryland, College Park, MD 20742, USA
   \and NASA Goddard Space Flight Center, Code 662, Greenbelt, MD 20771, USA
   \and RIKEN Cluster for Pioneering Research, 2-1 Hirosawa, Wako, Saitama 351-0198, Japan
   \and MIT Kavli Institute for Astrophysics and Space Research, Massachusetts Institute of Technology, Cambridge, MA 02139, USA
   \and INAF - Istituto di Astrofisica e Planetologia Spaziali, Via del Fosso del Cavaliere 100, 00133, Roma, Italy
   \and Yale Center for Astronomy \& Astrophysics, 52 Hillhouse Avenue, New Haven, CT 06511, USA
   \and Joint ALMA Observatory, Avenida Alonso de Cordova 3107, Vitacura 7630355, Santiago, Chile
   \and Eureka Scientific, 2452 Delmer Street Suite 100, Oakland, CA 94602-3017, USA
   \and Space Science Institute, 4750 Walnut Street, Suite 205, Boulder, CO 80301, USA
   \and Department of Physics and Astronomy, West Virginia University, P.O. Box 6315, Morgantown, WV 26506, USA
   \and Department of Astronomy, University of Maryland, College Park, MD 20742, USA
   \and Department of Astronomy, University of Geneva, 1205 Versoix, Switzerland
   \and National Radio Astronomy Observatory, 520 Edgemont Road, Charlottesville, VA 22903, USA
   \and Department of Astronomy, University of Florida, P.O. Box 112055, Gainesville, FL 32611, USA
   \and Department of Astronomy, University of Virginia, 530 McCormick Road, Charlottesville, VA 22904, USA
   \and School of Physics and Astronomy, Tel Aviv University, Tel Aviv 69978, Israel
    }

   \date{Received September 15, 1996; accepted March 16, 1997}
\abstract{We present the results of a joint \xmm and \nustar campaign on the active galactic nucleus (AGN) IC~4329A, consisting of 9 $\times$ 20\,ks \xmm observations, and 5 $\times$ 20\,ks \nustar observations within nine days, performed in August 2021. Within each observation, the AGN is not very variable, and the fractional variability never exceeds 5\%. Flux variations are observed between the different observations on timescales of days, with a ratio of 30\% of the minimum and maximum 2--10\,keV flux. These variations follow the softer-when-brighter behavior typically observed in AGN. In all observations, a soft excess is clearly present. Consistently with previous observations, the X-ray spectra of the source exhibit a cutoff energy between 140 and 250\,keV that is constant within the error in the different observations. We detected a prominent component of the $6.4$\,keV Fe~K$\alpha$ line consistent with being constant during the monitoring, consisting of an unresolved narrow core and a broader component likely originating in the inner accredion disk. We find that the reflection component is weak ($R_{\rm max}=0.009\pm0.002$) and most likely originates in distant neutral medium. We also found a warm absorber component together with an ultrafast outflow. Their energetics show that these outflows have enough mechanical power for significant feedback on the environment of the AGN.}
\maketitle
%
\nolinenumbers
\section{Introduction}
\label{sect:intro}
Active galactic nuclei (AGN) are extremely luminous compact objects located at the center of massive galaxies. They are powered by the accretion of gas onto the central supermassive black hole (SMBH; $M_{\rm BH } > 10^5 $\,$M_{\odot}$; \citealp{Salpeter1964,2008ARA&A..46..475H}). The X-ray emission of AGN is due to high-energy processes that occur in the so-called hot corona, which contains high-energy electrons and is located near the SMBH \citep{1991ApJ...380L..51H,1994ApJ...432L..95H,1997ApJ...476..620H,2003MNRAS.341.1051M,2009Natur.459..540F,2012MNRAS.422..129Z,2013MNRAS.431.2441D}. The hot electrons interact with thermal UV/optical photons emitted by the accretion disk through inverse-Compton scattering, which boosts the energy of the photons into the X-ray band. This process produces a broad power-law continuum in the X-ray spectrum with a cutoff at high energy (e.g., \citealt{1980A&A....86..121S}; \citealt{1993ApJ...413..507H}). Cold circumnuclear materials, such as the optically thick accretion disk and/or the molecular torus, absorb and reprocess the X-rays, which results in a feature in the X-ray spectrum that is known as Compton reflection, which peaks at around $\sim 30$\,keV \citep{1990Natur.344..132P,1991MNRAS.249..352G}. Additionally, the reflection process produces absorption and emission lines, such as the prominent Fe\ka emission line at 6.4\,keV \citep{1994MNRAS.268..405N}, which are the result of photoelectric absorption and fluorescence \citep{1997MNRAS.289..175M}.

IC~4329A is a bright nearby (z=0.01598, \citealt{Koss_2022}) AGN and it is classified in the optical as a Seyfert\,1.5 galaxy \citep{2022ApJS..261....4O}. It is located at the center of an edge-on host galaxy, with a dust lane passing through the nucleus. It also shows a dusty lukewarm absorber in the UV band (i.e., $T \sim 3\times 10^4$\,K, \citealt{2001ApJ...562L..29C}), characterized by saturated UV absorption lines (C IV, N V) near the systemic velocity of the host galaxy, which likely causes the reddening of the continuum and emission lines. The black hole mass of IC~4329A is $\log(M_{\rm BH})[M_{\odot}]=7.15^{+0.38}_{-0.26}$ \citep{2024arXiv240107676G}, and its Eddington ratio is $\lambda_{\rm Edd}=L_{\rm bol}/L_{\rm Edd}=0.17$. This was computed using the bolometric luminosities calculated from the intrinsic luminosities in the 14--150\,keV range, as shown in \citealt{Ricci2017}, with a bolometric correction of 8 \citep{2022ApJS..261....1K}. Assuming Equation 4 from \citet{Du2018}, the dimensionless accretion rate of IC~4329A is $\dot{\mathscr{M}}=1.4$ \citep{2024arXiv240107676G}.

IC~4329A has been extensively studied in the X-ray band in the past decades by all major X-rays satellites, including \textit{Einstein} \citep{1984ApJ...280..499P,1989ESASP.296.1105H}, EXOSAT \citep{1991ApJ...377..417S}, Ginga \citep{1990ApJ...360L..35P}, ASCA/RXTE \citep{2000ApJ...536..213D}, BeppoSAX \citep{2002A&A...389..802P,2007A&A...461.1209D}, \textit{Chandra} \citep{2004ApJ...608..157M,2010ApJS..187..581S}, \xmm \citep{2005A&A...432..453S,2007MNRAS.382..194N,2018A&A...619A..20M}, INTEGRAL \citep{2006ApJ...638..642B}, \textit{Swift}/BAT \citep{2009ApJ...690.1322W,Ricci2017}, \suzaku \citep{2014MNRAS.442L..95M} \nustar \citep{2014ApJ...781...83B,2014ApJ...788...61B}, and AstroSat \citep{2021ApJ...915...25T,2021MNRAS.504.4015D}. Its 2--10\,keV flux ranges between $8.1\times 10^{-11}$ \flux \citep{2009ApJ...690.1322W} and $1.8\times 10^{-9}$ \flux \citep{2010ApJS..187..581S}. Previous BeppoSAX, ASCA+RXTE INTEGRAL and \textit{Swift}/BAT observations of IC~4329A placed rough constraints on the high-energy cutoff at $E_{\rm c} \geq 180$\,keV \citep{2002A&A...389..802P}, $E_{\rm c} = 150-390$\,keV \citep{2000ApJ...536..213D}, $E_{\rm c}=60-300$\,keV \citep{2013MNRAS.433.1687M} and $E_{\rm c}>190$\,keV \citep{Ricci2017}, respectively. 
The \nustar hard X-ray spectrum of this AGN is characterized by a power law with a photon index of $\Gamma=1.73\pm0.01$ and a high-energy cutoff at $E_{\rm c}=186\pm14$\,keV \citep{2014ApJ...788...61B}. Previous analysis also showed a neutral reflection component with a reflection fraction of $R_{\rm refl}=0.42\pm0.02$ \citep{2014MNRAS.442L..95M}.

This source is well known for its strong Fe\,\ka line \citep{1990ApJ...360L..35P}. The Fe\,\ka line in IC~4329A shows a narrow core at $6.4$\,keV that is consistent with being produced by low-ionization material in the outer accretion disk, the broadline region (BLR), or in the torus. The narrow core component of the Fe\,\ka is also variable on timescales of weeks to months \citep{2016ApJ...821...15F,2022A&A...664A..46A}.
From the analysis of the ASCA, RXTE, BeppoSAX, \suzaku, and \nustar observations \citep{2000ApJ...536..213D,2007A&A...461.1209D,2014ApJ...788...61B,2014MNRAS.442L..95M}, a broad component was reported that is most likely produced in the inner part of the accretion disk and is blurred by general relativistic effects. In other works \citep[e.g.][]{2004ApJ...608..157M,2006ApJ...646..783M,2007MNRAS.382..194N,2021ApJ...915...25T}, the observations suggested a modest or weak broad iron line, indicating that X–ray reflection from the inner disk is weak in this source. Analyzing the \suzaku and \nustar observations of IC~4329A with a relativistic reflection model, \citet{2019ApJ...875..115O} inferred a very low reflection fraction ($R \sim 3.2 \times 10^{-3}$).

IC~4329A also shows an ionized warm absorber (WA) component with \logxi in the range of $-1.37$ and $2.7$ \citep{2005A&A...432..453S} and a highly ionized, that is, \logxi$=5.34\pm0.94$, ultrafast outflowing component (UFO; \citealt{2015Natur.519..436T}) with an outflow velocity $v/c=0.098\pm 0.004$. This component was first detected by \citet{2006ApJ...646..783M} and was then confirmed by \citet{2011ApJ...742...44T}.

Recently, IC~4329A was observed by the Imaging X-ray Polarimetry Explorer (IXPE) for $\sim 500$\,ks. From these observations, it appears that the source shows a $1\sigma$ confidence limit on a polarization degree of $3.3\%\pm1.1\%$ and a polarization angle of $78^{\circ}\pm10^{\circ}$, which is consistent with it being aligned with the radio jet \citep{2023MNRAS.525.5437I,2023JApA...44...87P}. These finding favor coronal geometries that are more asymmetric and might be outflowing. The coronal geometry is unconstrained within the $3\sigma$ confidence level, however.

We present the results of the analysis of the X-ray broadband spectra from the simultaneous \xmm and \nustar observations of IC~4329A. These observations are part of a joint millimeter/X-ray study of this nearby source, which also includes observations from \textit{Swift} (10 observations), NICER (20 observations), and from the Atacama Large Millimeter/submillimeter Array (\textit{ALMA}; 10 observations). The aim of the campaign is to study the relation between the 100\,GHz and X-ray continuum in AGN and to test the idea that the millimeter continuum is associated with self-absorbed synchrotron emission from the X-ray corona (e.g., \citealt{2008MNRAS.390..847L,2014PASJ...66L...8I}), as suggested by the recent discovery of a tight correlation between the X-ray luminosity and the 100\,GHz \citep{2023ApJ...952L..28R} and 200\,GHz \citep{2022ApJ...938...87K} luminosities of nearby AGN. We focus on the high-resolution X-ray spectroscopy of IC~4329A and therefore do not include \textit{Swift} and NICER data, which will be presented together with the results of ALMA campaign in a dedicated forthcoming paper (Shablovinskaya et al., in prep).

The paper is organized as follows: In Sect. \ref{sect:data_reduction}, we present the dataset we analyzed in this work. In Sect. \ref{sect:timing} and Sect. \ref{sect:spec}, we describe the timing and spectral data analysis processes, respectively. In Sect. \ref{sect:discussion}, we discuss the results of our analysis, which are summarized in Sect. \ref{sect:conclusion}. Standard cosmological parameters (H=70\,km\,s$^{-1} \rm Mpc^{-1}$, $\Omega_{\Lambda}$=0.73 and $\Omega_m$=0.27) are adopted.
\section{Data and data reduction}
\label{sect:data_reduction}
\begin{table}
\caption{Summary of the 2021 \xmm and \nustar observations of IC~4329A.}
\label{tab:observations_table}
\begin{tabular}{lccc}
\hline
\hline
Telescope & Obs. ID & Start Date & Exp.\\
 & & yyyy-mm-dd & ks\\
\hline
\hline
\xmm & 0862090101 & 2021-08-10 & 22.2\\
\nustar & 60702050002 & 2021-08-10 & 20.6\\
\xmm & 0862090201 & 2021-08-11 & 17\\
\xmm & 0862090301 & 2021-08-12 & 24\\
\nustar & 60702050004 & 2021-08-12 & 21.1\\
\xmm & 0862090401 & 2021-08-13 & 20\\
\xmm & 0862090501 & 2021-08-14 & 22.5\\
\nustar & 60702050006 & 2021-08-14 & 19.8\\
\xmm & 0862090601 & 2021-08-15 & 17.9\\
\xmm & 0862090701 & 2021-08-16 & 23\\
\nustar & 60702050008 & 2021-08-16 & 18.7\\
\xmm & 0862090801 & 2021-08-17 & 19.5\\
\xmm & 0862090901 & 2021-08-18 & 21.3\\
\nustar & 60702050010 & 2021-08-18 & 18\\
\xmm & 0862091001 & 2021-08-19 & 21.3\\
\hline
\hline
\end{tabular}
\end{table}
The dataset analyzed in this work consists of nine X-ray Multi-Mirror Mission ({\it XMM-Newton}; \citealt{Jansen2001}) observations, five of which were performed simultaneously with the Nuclear Spectroscopic Telescope Array ({\it NuSTAR}; \citealt{Harrison2013}). Details of the duration and exposure of the observations are reported in Table \ref{tab:observations_table}.

\subsection{\xmm data reduction}
IC~4329A was observed once per day for ten consecutive days (from 10 August 2021 to 19 August 2021) by \xmm (P.I. C. Ricci) during the \xmm AO\,19.
The \xmm observations were performed with the European Photon Imaging Camera (EPIC hereafter) detectors, and with the Reflection Grating Spectrometer (RGS; \citealt{denHerder2001}). The EPIC cameras were operated in small-window and thin-filter mode. Observation 0862090401 is not included in the analysis because due to a problem in the ground segment, the EPIC-pn exposure was lost.

The EPIC camera \citep{Struder2001} event lists were extracted with the \textsc{epproc} and \textsc{emproc} tools of the standard system analysis software (\textsc{SAS} v.18.0.0; \citealt{Gabriel2004}). The MOS detectors \citep{Turner2001} and the RGS were not considered because due to the lower statistics of their spectra, they do not add information to the analysis.

The choice of the optimal time cuts for the flaring particle background and of the source and background extraction radii was performed via an iterative process that maximized the signal-to-noise ratio (S/N) as in \citet{Piconcelli2004}. The resulting optimum extraction radius was $30\arcsec$, and the background spectra were extracted from source-free circular regions with radii of $\sim$ $50\arcsec$ for EPIC-pn for each observation. Response matrices and auxiliary response files were generated using the SAS tools \textsc{rmfgen} and \textsc{arfgen}, respectively. EPIC-pn spectra were binned in order to oversample the instrumental energy resolution by a factor larger than three and to have no fewer than 20 counts in each background-subtracted spectral channel. No significant pile-up affected the EPIC data, as indicated by the SAS task \textsc{epatplot}. EPIC-pn light curves were extracted by using the same circular regions for the source and the background as the spectra.

\subsection{\nustar data reduction}
The five \nustar observations were performed simultaneously with five \xmm observations during \nustar AO\,7 (P.I. C. Ricci). The \nustar telescope observed the source with its two coaligned X-ray telescopes Focal Plane Modules A and B (FPMA and FPMB, respectively).

The \nustar high-level products were obtained using the \nustar data analysis software (\textsc{NuSTARDAS}) package (v2.1.1) within the \textsc{heasoft} package (version 6.29). Cleaned event files (level 2 data products) were produced and calibrated using standard filtering criteria with the \textsc{nupipeline} task and the latest calibration files available in the \nustar calibration database (CALDB 20220802). For both FPMA and FPMB, the radii of the circular region used to extract source and background spectra were $75\arcsec$; no other bright X-ray source is present within $75\arcsec$ from IC~4329A, and no source was present in the background region. The low-energy (0.2--5\,keV) effective area issue for FPMA \citep{Madsen2020} does not affect our observations because no low-energy excess is found in the spectrum of this detector. The spectra were binned in order to oversample the instrumental resolution by a factor of 2.5 at least and to have an S/N higher than 3 in each spectral channel. Light curves were extracted using the \textsc{nuproducts} task, adopting the same circular regions as for the spectra. 
\section{Timing analysis}
\label{sect:timing}
\begin{figure*}
        \includegraphics[width=\textwidth]{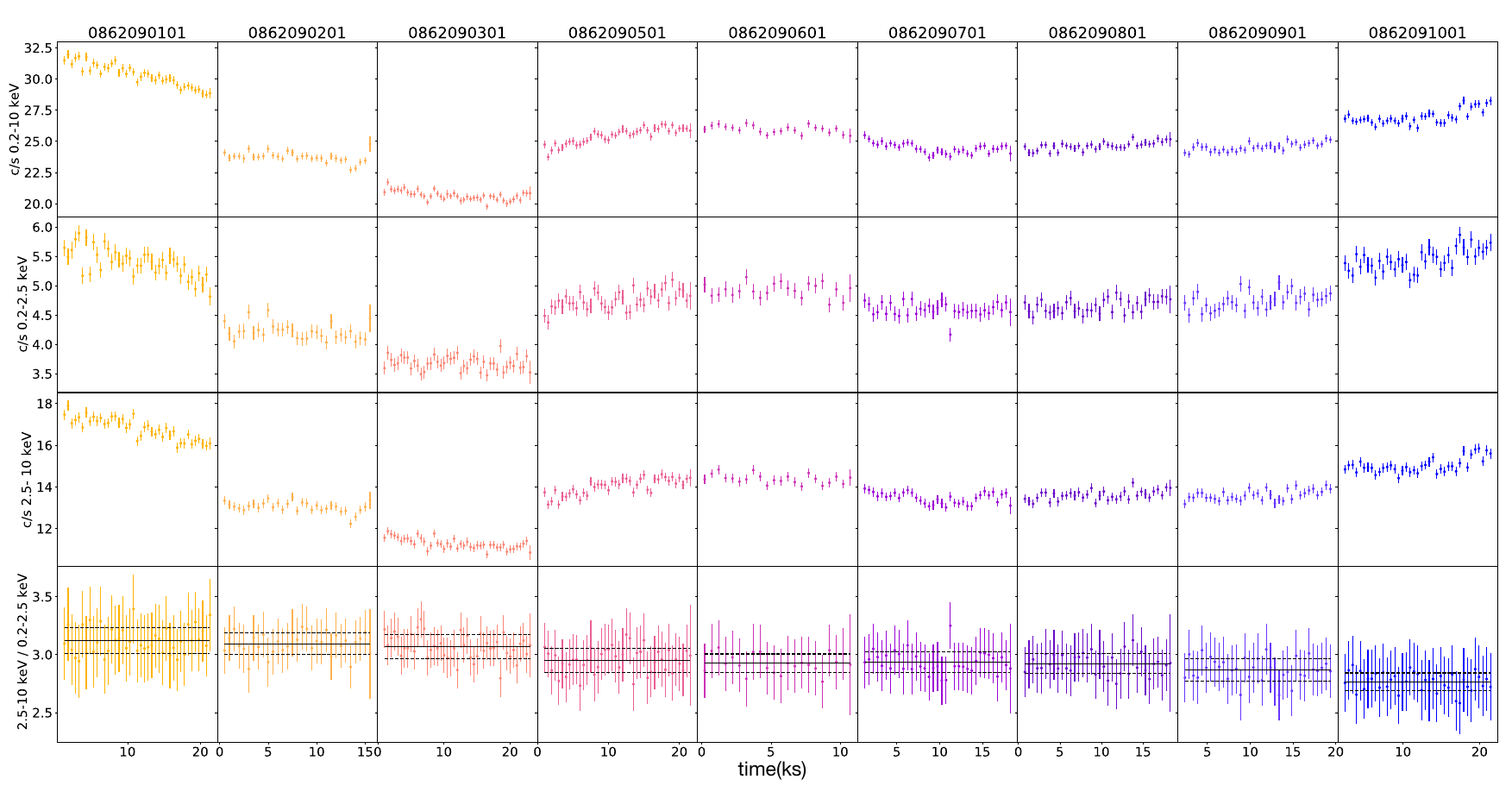}
    \caption{\xmm EPIC-pn light curves (background subtracted) in the 0.2--10\,keV, 0.2--2.5\,keV and in the 2.5--10\,keV are shown in the top, top middle, and bottom middle panels, respectively. In the bottom panels, the ratios of the \xmm EPIC-pn background-subtracted light curves in the 2.5--10\,keV and 0.2--2.5\,keV energy bands are reported.  All the light curves are extracted using a binning time of 500 s. The solid and dashed black lines indicate the average and the standard error of the mean, respectively. The color code used in this figure for different ObsID is applied throughout the paper.}
    \label{fig:xmm_lc}
\end{figure*}
\begin{figure*}
        \includegraphics[width=\textwidth]{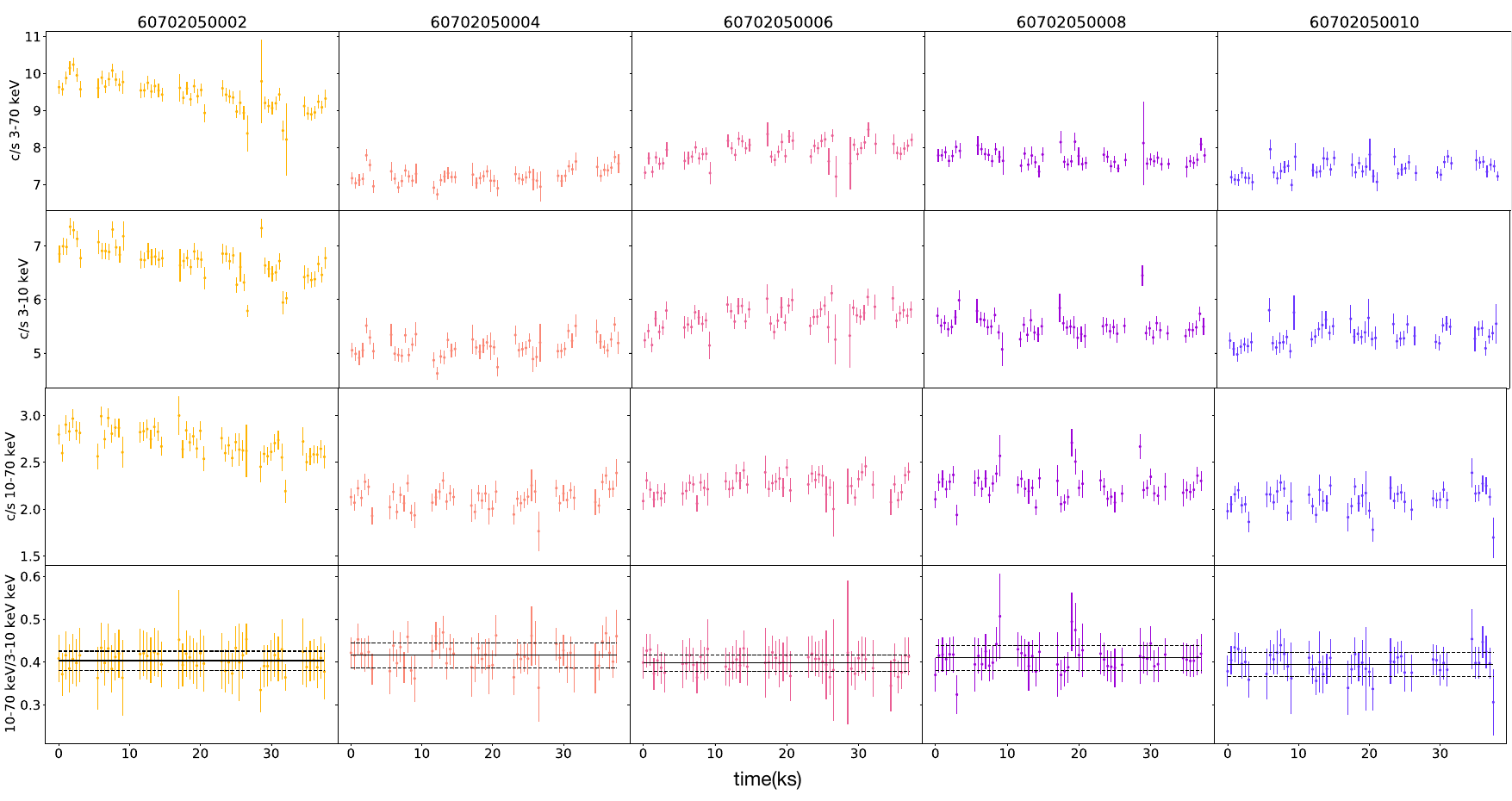}
    \caption{\nustar FPMA+B background-subtracted light curve in the 3--70\,keV. 3--10\,keV and 10--70\,keV energy ranges are shown in the top, top middle, and bottom middle panels, respectively. In the bottom panels, the ratios of the \nustar NuSTAR background-subtracted light curves in the 10--70\,keV and 3--10\,keV energy bands are reported. All the light curves are extracted using a binning time of 500\,s. The color code used in this figure for different ObsID is applied throughout the paper. The solid and dashed black lines indicate the average and the standard error of the mean, respectively.}
    \label{fig:nu_lc}
\end{figure*}
During the monitoring, IC~4329A varied within some observations (see Fig.\,\ref{fig:xmm_lc} and \ref{fig:nu_lc}). However, these variations did not affect the ratio of the flux in the 2.5--10\,keV and the 0.2--2.5\,keV band, (< 15\%; see the bottom panels of Fig. \ref{fig:xmm_lc} and \ref{fig:nu_lc}).

We studied the variability spectra of the \xmm EPIC-pn observations of IC~4329A using one of the common estimators for X-ray variability: the fractional variability F$_{\rm var}$ \citep{2004MNRAS.351..193V,2004MNRAS.348..207P,2006Natur.444..730M,2012A&A...542A..83P,2016MNRAS.458.1311M,2017MNRAS.465.2804M,2020A&A...634A..65D}. The F$_{\rm{var}}$, that is, the square root of the normalized excess variance ($\sigma^2_{\rm NXS}$ \citealt{1997ApJ...476...70N,2002ApJ...568..610E,2003MNRAS.345.1271V}), is the difference between the total variance of the light curve and the mean squared error that is normalized for the average of the number of flux measurements squared. N is the number of good time intervals in a light curve, and $x_i$ and $\sigma_i$ are the flux and the error in each interval. F$_{\rm{var}}$ is defined \citep{2003MNRAS.345.1271V} as
\begin{equation}
F_{\rm var}=\sqrt{\sigma^2_{\rm NXS}}=\sqrt{\frac{S^2-\overline{\sigma^2}}{\overline{x_i}^2}},
\label{eq:excess_var}
\end{equation}
where
$S^2=\frac{1}{N-1}\sum_{i=1}^{N}[(x_i-\overline{x_i})^2]$ is the sample variance, that is, the integral of the power spectral density (PDS) between two frequencies, and $ \overline{\sigma^2}=\frac{1}{N}\sum_{i=1}^{N}[\sigma_i^2]$ is the mean square error.

\begin{figure}
\includegraphics[width=0.45\textwidth]{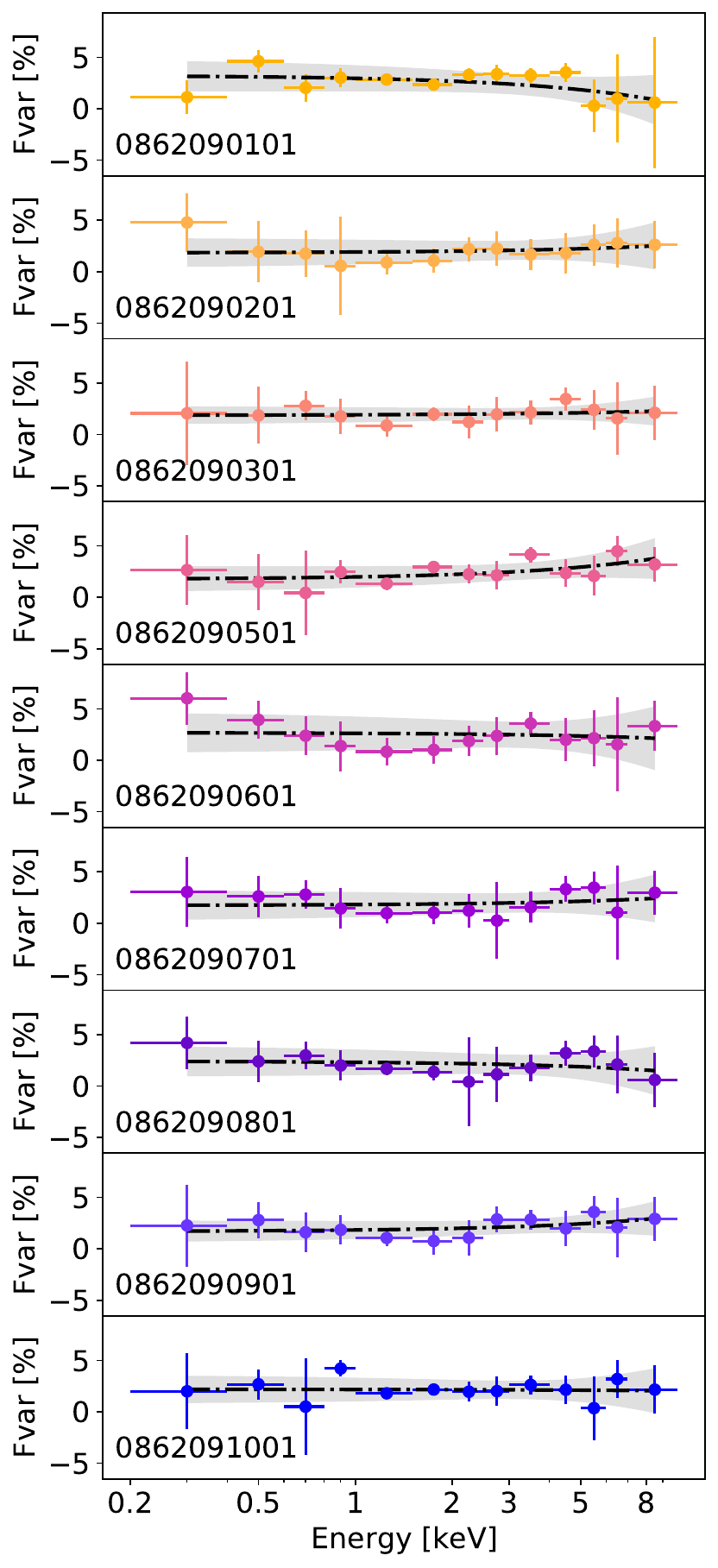}
\caption{Fractional variability spectra for the \xmm EPIC-pn observations of IC~4329A as in Table\,\ref{tab:observations_table}. The spectra of this campaign are characterized by a mostly flat shape, with a fractional variability that never exceeds 5\%. The dot-dashed black lines are the linear regressions, and the shaded regions represent the combined 3$\sigma$ error on the slope and normalization.}
\label{fig:fvar}
\end{figure}
\begin{figure}
\includegraphics[width=0.5\textwidth]{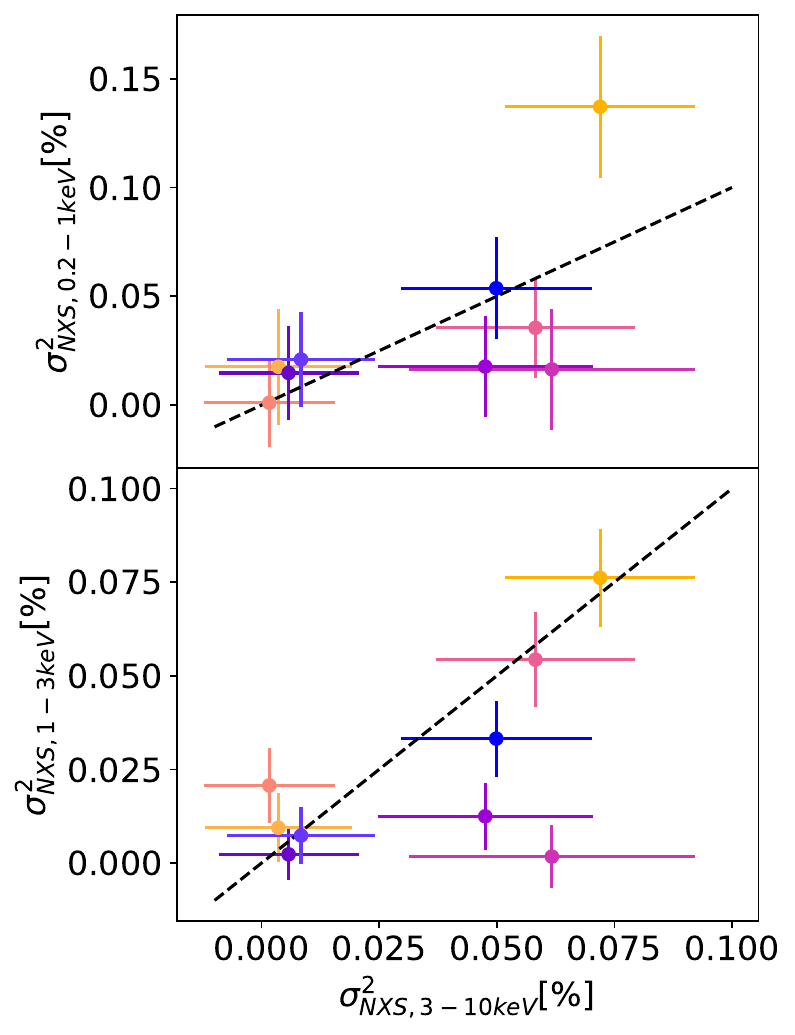}
 \caption{Top panel: Soft (0.2--1\,keV) vs. hard (3--10\,keV) $\sigma_{\rm NXS}^2$. Bottom panel: Medium(1--3\,keV) vs. hard (3--10\,keV) $\sigma_{\rm NXS}^2$. The values are calculated for the \xmm EPIC-pn observations of IC~4329A as in Table\,\ref{tab:observations_table}. The dashed black lines represent the one-to-one relation. The color code for the different observations is the same as in Fig. \ref{fig:xmm_lc}.} \label{fig:nxs}
\end{figure}
We computed F$_{\rm var}$ for each observation using the background-subtracted EPIC-pn light curves in different energy bands using a temporal bin of 1000\,s on a timescale of 20\,ks. The errors were computed using Eq.\,B2 of \citet{2003MNRAS.345.1271V}. The resulting F$_{\rm var}$ spectra of each observation are shown in Fig. \ref{fig:fvar}. To study the behavior of the variability spectra, we used the \texttt{linmix} code, a hierarchical Bayesian model for fitting a straight line to data with errors in both the x and y directions \citep{Kelly_2007}. To perform the fitting, we used a linear model of the data with the following fitting relation:
\begin{equation}
    F_{\rm var}=\alpha \times E+\beta
    \label{eq:Fvar_fit}.
\end{equation}
We found the behavior of the variability spectra of this campaign to be nearly flat, with a correlation coefficient $\rho_{\rm Pearson}$ spanning between -0.05 and 0.05.

The different spectral components observed in AGN can lead to spectral variability in different energy bands as each of them can dominate in a certain energy band. For example, the primary power-law component or the reflection component are dominant in the hard energy band (3--10\,keV, \citealt{1991ApJ...380L..51H,1993ApJ...413..507H,1993MNRAS.261..346H}), while soft-excess and WA can dominate the soft (0.2--1\,keV, \citealt{Bianchi2009}) and medium (1--3\,keV, \citealt{Blustin2005,Tombesi2013}) energy bands. We therefore calculated the $\sigma_{\rm NXS}^2$ from the 0.2--1\,keV (soft), 1--3\,keV (medium), and 3--10\,keV (hard) light curves to obtain a complete picture of the X-ray variability of IC~4329A, and we compare these values in Fig. \ref{fig:nxs}. The $\sigma_{\rm NXS}^2$ in the soft, medium, and hard energy band for the \xmm observations of IC~4329A in this campaign are very low in generla. They are below 0.1\% in all the observations, except for the $\sigma^2_{\rm NXS}$ in the soft energy band of Obs.ID 0862090101 (0.15\%). The $\sigma^2_{\rm NXS, 1-3\,keV}$ vs $\sigma^2_{\rm NXS, 3-10\,keV}$ values are located in the vicinity or slightly below the one-to-one relation. The same holds for the $\sigma^2_{\rm NXS, 0.2-1\,keV}$ vs $\sigma^2_{\rm NXS, 3-10\,keV}$ values, except again for Obs.ID 0862090101, which is located above the one-to-one relation. This could also be related to the small peak of the variability spectrum of this observation at around 0.5\,keV. Thus, IC~4329A shows very weak variations overall, and  the soft-excess and/or WA component variations are even weaker than those of the primary continuum and/or reflection component on a timescale of 20\,ks (except for Obs.ID 0862090101, in which most likely the soft-excess or the absorbing components are more variable on timescales shorter than $\sim 20$\,ks with respect to the continuum), in agreement with what is expected for type 1 AGN \citep{2012A&A...542A..83P,2023MNRAS.526.1687T}.

The \nustar light curves show no spectral variations($<10$\,\%) in the ratio of the 3--10 and 10--70\,keV count rates (see the bottom panels of Fig. \ref{fig:nu_lc}).

Since the hardness ratios of the \xmm and \nustar observations do not change strongly within each pointing (see the bottom panels of Fig.\,\ref{fig:xmm_lc} and \ref{fig:nu_lc}), we decided to use the time-averaged spectrum for our spectral analysis to improve the spectrum statistics.
\section{Spectral analysis}
\label{sect:spec}
\begin{figure}
    \includegraphics[scale=0.35]{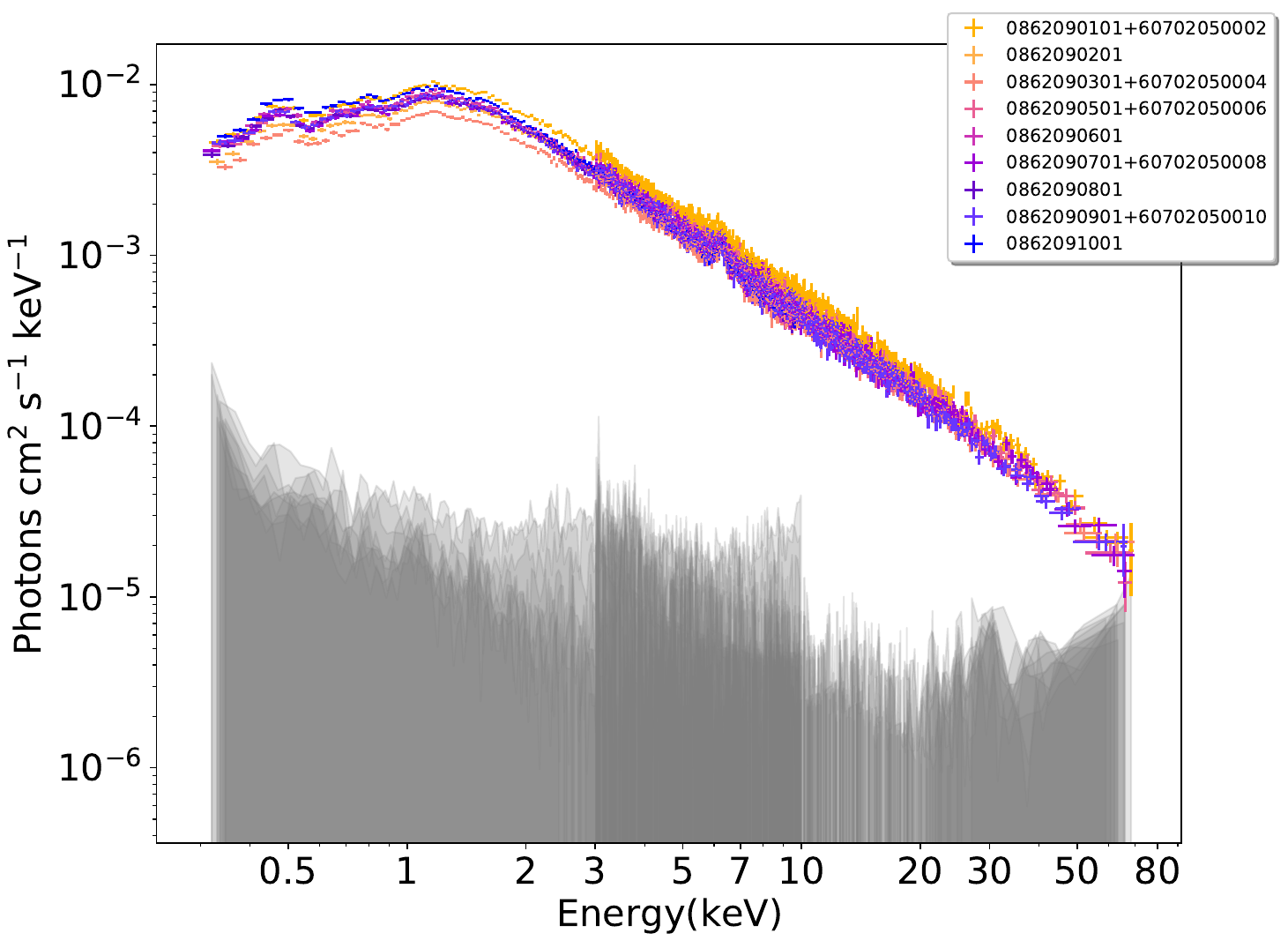}
    \caption{Background-subtracted \xmm EPIC-pn and \nustar FPMA-B spectra and the corresponding background levels (shaded regions). The simultaneous \xmm and \nustar observations are reported with the same color. The color code for different observations is the same as in Fig. \ref{fig:xmm_lc}. This color code is applied throughout the paper.}
    \label{fig:data_bkg}
\end{figure}
We performed the spectral analysis using the \textsc{xspec} v.12.12.1 software package \citep{Arnaud1996}. The spectra obtained by the two \nustar modules (FPMA and FPMB) were fit simultaneously, with a typical cross-normalization constant lower than 5\% \citep{Madsen2015}.

Errors and upper and/or lower limits were calculated using the $\Delta\chi^2$ = 2.71 criterion (corresponding to the 90\% confidence level for one parameter), unless stated otherwise.

The Galactic column density at the position of the source ($N_{\rm H}=4.1 \times 10^{20} \rm cm^{-2}$; \citealp{HI4PI2016}) was always included in the fits, and it was modeled with the \textsc{tbabs} component \citep{Wilms2000} in \textsc{xspec}, with $N_{\rm H}$ kept frozen to the quoted value. We also assumed solar abundances unless stated otherwise.

Fig. \ref{fig:data_bkg} shows the background-subtracted EPIC pn (0.3--10\,keV), and FPMA-B spectra (3--75\,keV), plotted with the corresponding X-ray background. In all figures, the spectra were corrected for the effective area of each detector.
\subsection{\xmm data analysis}
\label{sect:xmm}
We started our spectral analysis with the 2022 EPIC-pn \xmm spectra (in the 0.3--10\,keV range). We fit the spectra separately and adopted a fitting model that was based on literature results. The model is composed of a neutral absorption component at the redshift of the source, a WA component \citep{2005A&A...432..453S}, and an UFO component \citep{2006ApJ...646..783M,2011ApJ...742...44T}, \textsc{xspec} model: \textsc{tbabs *  ztbabs *  mtable(WAcomp) *  mtable(UFOcomp) *  (bbody +  powerlaw + zgauss + zgauss), hereafter Model\,XMM}. The two latter components were modeled with detailed grids computed using the photoionization code \textsc{xstar} \citep{Kallman2001}, with an input spectral energy distribution described by a power law with a photon index of $\Gamma = 2$. These tables consider standard solar abundances from \citet{Asplund2009}, and take into account absorption lines and edges for all the metals characterized by an atomic number $Z \leq 30$. Considering the typical values of turbulent velocity for WAs \citep{Laha2014}, we used a table computed considering a turbulent velocity of 100\,$\rm km\,s^{-1}$ for the WA and a table computed considering a turbulent velocity of 1000\,$\rm km\,s^{-1}$ for the UFO. We added a blackbody component to take into account any weak underlying soft X-ray spectral feature, such as the soft excess. The primary continuum was modeled by a simple power law. We focused on the \xmm energy band range and therefore did not include reflection features at this stage. Two Gaussian lines were included to represent the iron \ka and \kb lines. The line widths of both lines were free to vary, as well as their centroid energies. The fits obtained using this model are very good (see the left panels of Fig. \ref{fig:ratios}). The values of the Fe \ka line parameters are shown in Table\,\ref{tab:kalpha}. The line width of the \ka line is consistent within the errors with the \xmm resolution $\sim 200$\,eV. The line width of the Fe\,\kb line is $\sim80-90$\,eV because it fits a blend of lines that might also include \Fexxvi. The line width of the Fe\,\ka line in some observations is unconstrained, and we found only upper limits. When we tried to include a narrower core for the Fe\,\ka, by adding another \textsc{zgauss} component with a line width fixed to 1\,eV, we were not able to recover the line flux for both components. We either found upper limits for the narrower or for the broader component. This suggests that if a broad component of the Fe \ka line and a narrow core are present, they are unresolved during our X-ray monitoring campaign. To determine whether the broad unresolved Fe \ka component is due to relativistic reflection off the innermost regions of the accretion disk, we replaced the Gaussian component with the \textsc{diskline} model. From this analysis, we found that a subdominant ($EW\sim15$\,eV) relativistically broadened Fe \ka component it is required by the data in 50\% of the observations of the monitoring, where we found an inner radius $R_{\rm in}\sim 7 R_{\rm g}$ at which $R_{\rm g}=GM_{\rm BH}/c^2$. \\
\begin{table}
\centering
\caption{Details of the iron \ka line parameters from the analysis of the \xmm observations. The flux of the Fe \ka line is given in units of \flux.}
\label{tab:kalpha}
\begin{tabular}{lcccc}
\hline \hline
Obs.ID & E[keV] & $\sigma$[eV] & EW [eV] & $\log(F)$\\
\hline
\hline
0862090101&$6.39\pm0.03$& $<92$             & $64\pm12$  & $-12.12\pm0.08$\\
0862090201&$6.36\pm0.02$& $65_{-32}^{+99}$  & $96\pm15$  & $-12.03\pm0.08$\\
0862090301&$6.36\pm0.01$& $93_{-17}^{+132}$ & $112\pm18$ & $-12.02\pm0.07$\\
0862090501&$6.37\pm0.02$& $<82$             & $46\pm 9$  & $-12.30\pm0.08$\\
0862090601&$6.36\pm0.02$& $81_{-42}^{+122}$ & $115\pm23$ & $-11.97\pm0.10$\\
0862090701&$6.38\pm0.02$& $<48$             & $81\pm19$  & $-12.09\pm0.12$\\
0862090801&$6.38\pm0.02$& $87_{-54}^{+114}$ &$104\pm26$  & $-12.01\pm0.07$\\
0862090901&$6.35\pm0.03$& $69_{-35}^{+132}$ & $78\pm29$  & $-12.14\pm0.11$\\
0862091001&$6.39\pm0.02$& $99_{-66}^{+131}$ & $108\pm31$ & $-11.99\pm0.07$\\
\hline
\hline
\end{tabular}
\end{table}
In the soft energy band (0.3--2\,keV), we found a moderate soft-excess component with temperatures ranging from kT$_{\rm BB}=47\pm 15$\,eV to kT$_{\rm BB}=75\pm 6$\,eV and a neutral absorption component at the redshift of the source with $N_{\rm H}=(2.67\pm0.07) \times 10^{21}\,$cm$^{-2}$ over the 9 \xmm observations. We also confirmed the presence of two X-ray ionized absorbing components, consistent with previous studies \citep{2005A&A...432..453S,2006ApJ...646..783M,2011ApJ...742...44T}. These components are constant within their uncertainties during the monitoring. We found for the X-ray wind with turbulent velocity 100\,$\rm km\, s^{-1}$, identified with a WA, an average column density $N_{\rm H} =(1.49\pm0.16) \times 10^{21}\,$cm$^{-2}$ , an average ionization fraction $\log(\xi  / \rm erg\,\rm s^{-1} \rm cm)=1.30\pm0.07$, and an average observed redshift of the absorption components z=$-0.011\pm0.005$. The component with a turbulent velocity 1000\,$\rm km\, s^{-1}$, identified with a UFO, shows an average column density $N_{\rm H} =(1.97\pm0.82) \times 10^{21}\,$cm$^{-2}$, an average ionization fraction $\log(\xi  / \rm erg\,\rm s^{-1} \rm cm)=2.69\pm0.03$, and an average  z=$-0.177\pm0.016$.
\subsection{\nustar data analysis}
\label{sect:nustar}
\begin{figure*}
\includegraphics[width=0.3\textwidth]{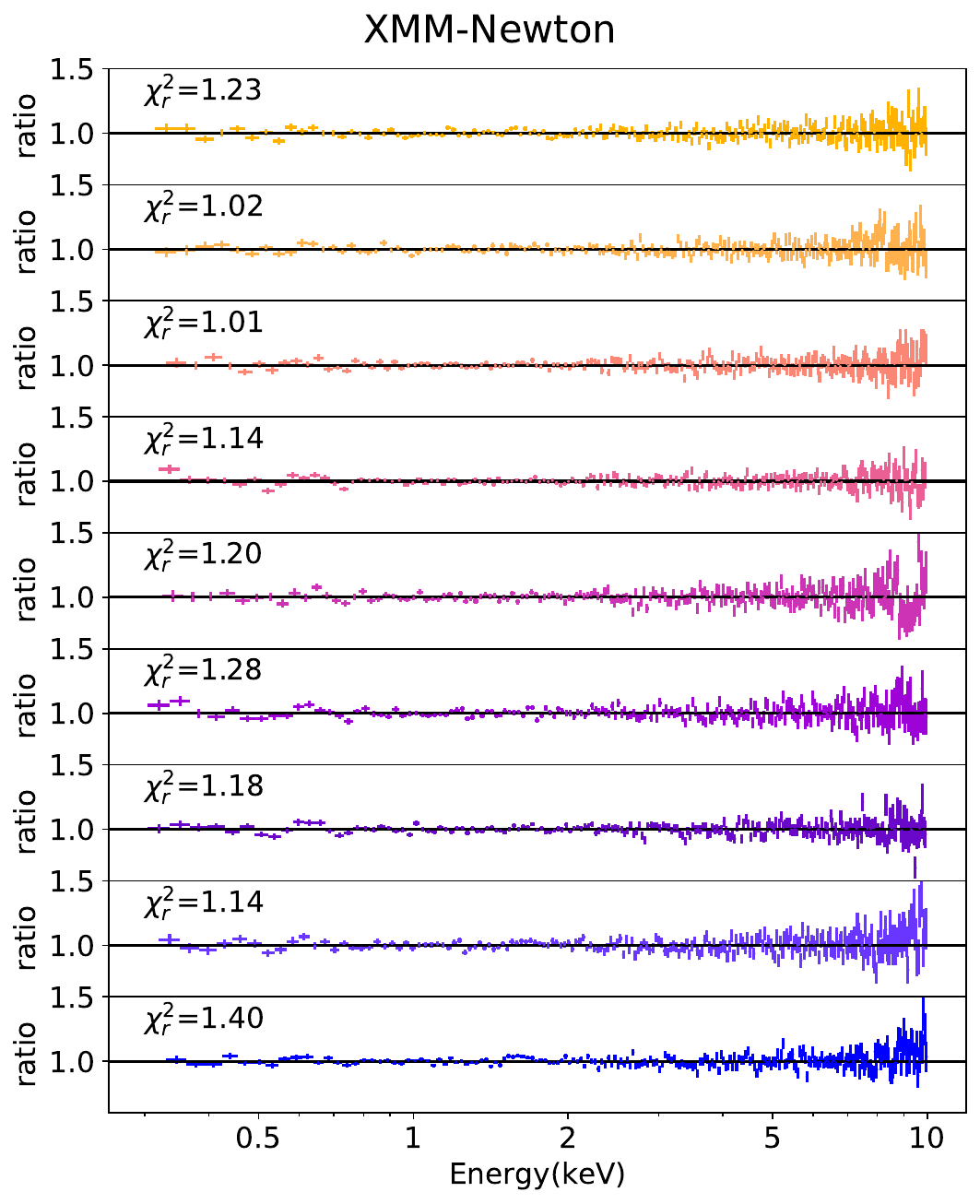}
\includegraphics[width=0.3\textwidth]{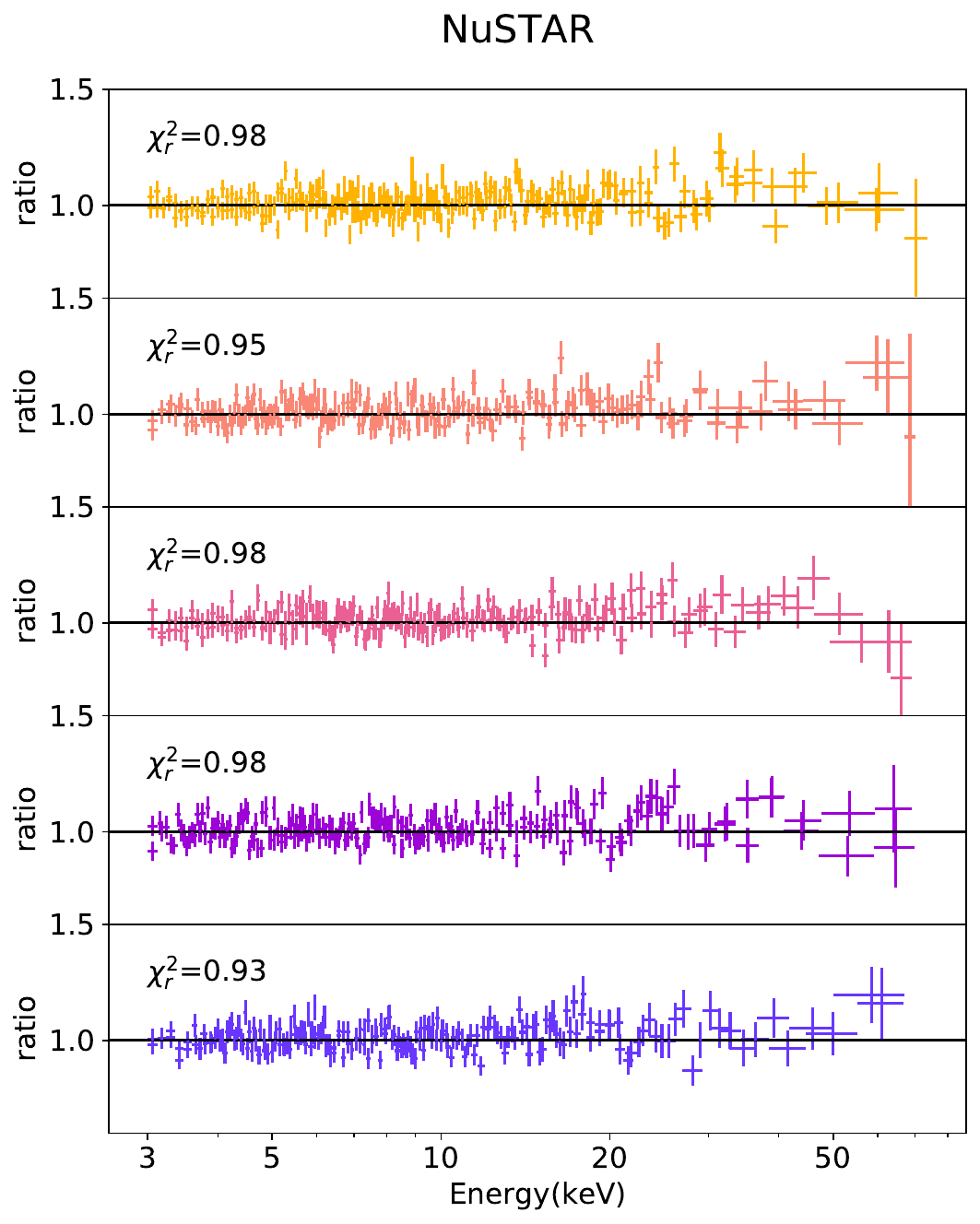}
\includegraphics[width=0.3\textwidth]{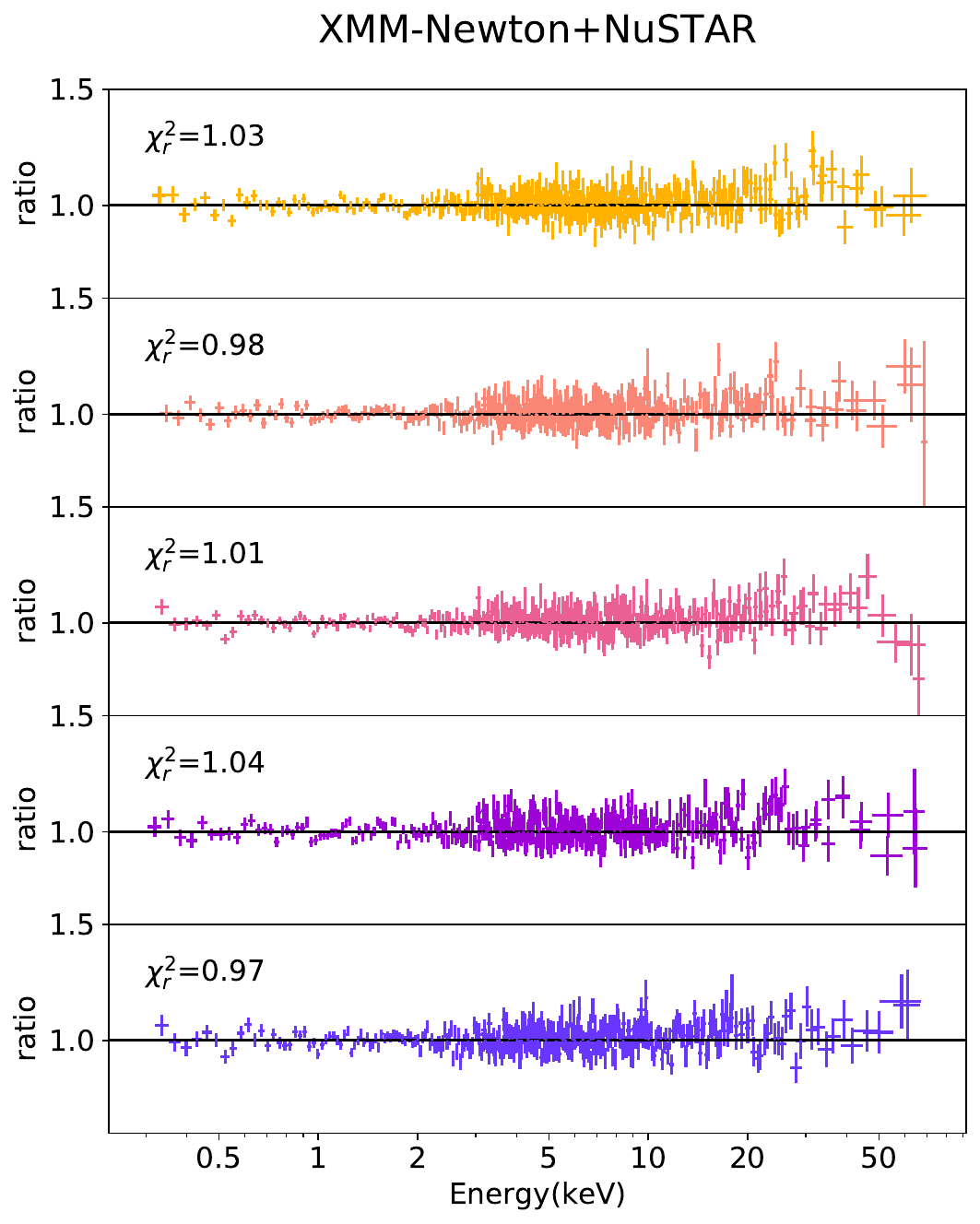}
\caption{Data/model ratio for the EPIC-pn \xmm observations (left panels), FPMA+B \nustar observations (central panels), and the X-ray broadband observations (right panels) of IC~4329A when Model\,XMM, Model\,Nu and Model\,A are applied, respectively (see Sect. \ref{sect:spec}).}
\label{fig:ratios}
\end{figure*}

Before analyzing the broadband spectra, we first focused on the primary emission and on its reflected component using \nustar FPMA+B spectra (in the 3--75\,keV range). During the fitting process, we left the \textit{NuSTAR} FPMB cross-calibration constant free to vary. Again, we composed a model following the models presented in the literature (e.g., \citealt{2014ApJ...781...83B,2014ApJ...788...61B}). We modeled the primary X-ray continuum with the \textsc{cutoffpl} model in \textsc{xspec}, which includes a power law with a high-energy exponential cutoff, and the reprocessed emission using the standard neutral reflection model \textsc{xillver} version [1.4.3] with $\log(\xi)$ fixed to zero \citep{Garcia2013} (\textsc{xspec} model: tbabs*(cutoffpl+ xillver), hereafter Model\,Nu). The cutoff energy, $E_{\rm c}$, and the photon index, $\Gamma$, of the \textsc{xillver} component were linked those of the \textsc{cutoffpl}, and the reflection fraction, $R_{\rm refl}$, was forced to be negative. With these settings, \textsc{xillver} only reproduced the reflection component. The disk inclination angle was fixed to a value of $60^{\circ}$ \citep{2014ApJ...788...61B}. The data are very well reproduced by the model. The ratio of data and model are shown in the central panels of Fig. \ref{fig:ratios}, in which we also report the $\chi^2_r$ = $\chi^2$/degrees of freedom (dof). In this analysis, we found a photon index of the primary power law ranging from $\Gamma=1.67\pm 0.04$ to $\Gamma=1.74 \pm 0.03$ and a cutoff value $E_{\rm c}= 145 - 180$\,keV, in agreement with previous studies (e.g., \citealp{2014ApJ...788...61B}). However, with respect to the results obtained by \citet{2014ApJ...788...61B}, we found a significantly lower value for the reflection fraction, R$_{\rm refl}=0.008 - 0.01$, similar to the value found in the analysis of \suzaku and \nustar observations of IC~4329A by \citet{2019ApJ...875..115O}. Together with the result of our analysis of the Fe \ka line with the \xmm observations, which shows faint broad features (see Sect. \ref{sect:xmm}), this suggests that the reflection from the inner disk is weak.

We also tested a scenario in which the Fe \ka emission is slightly broadened by relativistic effects and comes from the accretion disk by substituting the \textsc{xillver} component with the \textsc{relxill} component \citep{Garcia2014, Dauser2014}, which takes ionized relativistic reflection from an accretion disk illuminated by a hot corona into account. This model resulted in a significantly poorer fit ($\chi^2_r \sim 1.4 - 1.7$). Moreover, we were not able to constrain the black hole spin in any of the \nustar observations of this campaign, finding just a lower limit of $a \gtrsim 0.1$.
\subsection{Broadband data analysis}
\label{sect:broadband}
\begin{figure}
\centering
        \includegraphics[width=0.9\columnwidth]{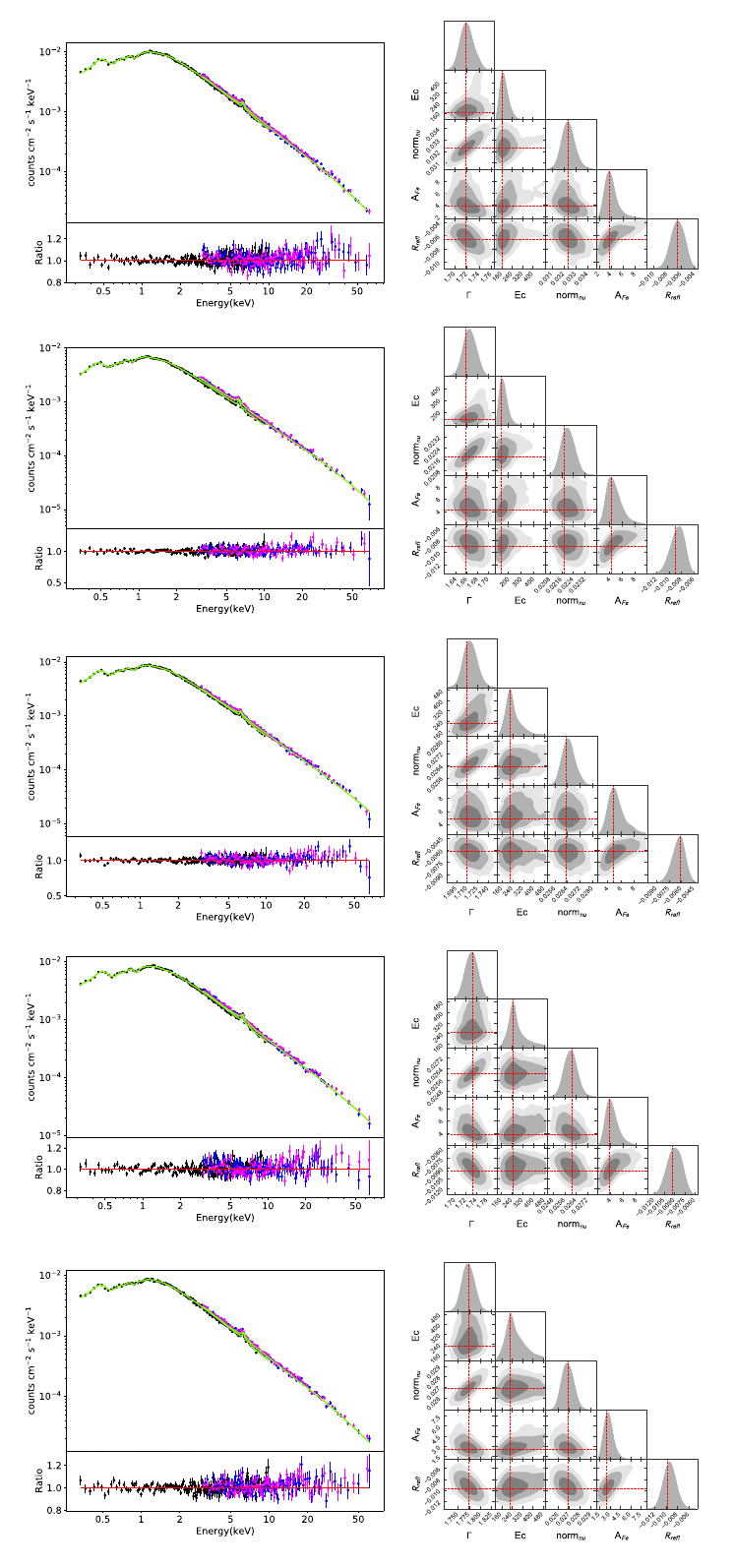}
    \caption{Left panels: Data, fitting model, and residual ratio for \xmm EPIC-pn (black) and \nustar FPMA (blue) and FPMB (magenta) spectra of the simultaneous observations of IC~4329A when Model\,A (see Sect. \ref{sect:broadband}) is applied to the data. Right panels: Contour plots at 68\%, 90\%, and 99\% resulting from the MCMC analysis of Model\,A applied to the broadband \xmm and \nustar 0.3--75\,keV spectra of IC~4329A. We show the outputs for the photon index ($\Gamma$), cutoff energy ($E_{\mathrm{cut}}$[keV]), \nustar normalization (norm$_{\rm nu}$ [ph/keV/cm$^2$/s]), iron abundance ($A_{\rm Fe}$), and reflection fraction ($R_{\rm refl}$). We report from top to bottom the Obs.IDs 086209101+60702050002, 086209103+60702050004, 086209105+60702050006, 086209107+60702050008, and 086209109+60702050010.}
    \label{fig:cont_spec}
\end{figure}

\begin{table*}
\centering
\caption{Best-fitting parameters for the X-ray broadband \xmm plus \nustar spectra of IC~4329A. Errors are at the 90\% confidence levels.}
\label{tab:bb_fit}
\begin{tabular}{llccccc}
\hline
\hline
Model & Parameter & \multicolumn{5}{c}{Obs.ID} \\
 \hline
 & & 0862090101 & 0862090301 & 0862090501 & 0862090701 & 0862090901\\
 & & 60702050002 & 60702050004 & 60702050006 & 60702050008 & 60702050010\\
\hline
Model\,A &$\frac{\chi^2}{ \rm dof} = \chi^2_r$ & 1.03 & 0.98 &1.01 & 1.04 &0.97 \\
\hline
 \textsc{cutoffpl}& $\Gamma$ & $1.71\pm 0.01$ & $1.65\pm0.02$ & $1.71\pm0.02$ & $1.73\pm0.02$ & $1.77\pm0.02$\\
 \textsc{cutoffpl}&$E_{\mathrm{cut}}$(keV) &$163^{+55}_{-35}$ & $140^{+56}_{-75}$ & $220^{+89}_{-45}$& $249^{+70}_{-92}$ & $228^{+57}_{-41}$\\
 \textsc{xillver}&R $_{\mathrm{refl}}$ & $0.006\pm 0.001$ & $0.008\pm0.002$ & $0.006\pm0.001$ & $0.009\pm0.002$& $0.009\pm0.001$\\
 \textsc{xillver}&A $_{\mathrm{Fe}}$  & $3.41^{+1.09}_{-0.89} $ & $3.72^{+2.46}_{-1.31}$& $3.77^{+1.30}_{-1.17}$ & $3.1^{+1.28}_{-0.53}$& $1.91^{+1.24}_{-0.81}$\\
\hline
Model\,B1 &$\frac{\chi^2}{ \rm dof} = \chi^2_r$ & 1.06 & 1.002 & 1.03 & 1.06 & 1.005  \\
\hline
\textsc{cutoffpl}& $\Gamma$ & $1.75\pm 0.02$ & $1.71\pm0.01$ & $1.73\pm 0.01$ & $1.74\pm0.02$ & $1.81\pm0.02$\\
\textsc{cutoffpl}&$E_{\mathrm{cut}}$(keV) &$>214$ &$>171$&$>210$& $>245$ & $>305$\\
\textsc{mytorus}&$N_{\rm H,tor} [10^{23}\rm cm^{-2}]$& $2.84\pm0.63$ & $2.78\pm0.95$ & $2.67\pm0.48$ &$2.32\pm0.84$  &$2.48\pm0.65$\\
\hline
Model\,B2 &$\frac{\chi^2}{ \rm dof} = \chi^2_r$ & 1.07 & 1.01 & 1.06 & 1.11 & 1.006 \\
\hline
\textsc{RXTorusD}& $\Gamma$ & $1.73\pm 0.01$ & $1.70\pm0.02$ & $1.73\pm0.04$ & $1.76\pm0.01$ & $1.76 \pm 0.01$\\
\textsc{RXTorusD}&$N_{\rm H,tor} [10^{23}\rm cm^{-2}]$& $2.13^{+0.83}_{-0.38}$& $3.70 \pm 0.51$ & $2.34\pm0.36$ & $5.10\pm0.56$&$3.57\pm0.76$\\
\textsc{RXTorusD}&$i({\rm deg})$& $44.91\pm0.98$& $44.66\pm 1.25$& $44.04 \pm 1.66$ & $44.83\pm0.77$ &$45.00\pm0.34$\\
\textsc{RXTorusD}&$r/R$& $0.69\pm0.02$& $0.69 \pm 0.01$ & $0.69\pm 0.02$ & $0.69 \pm 0.01$ &$0.69 \pm 0.03$\\
\hline
Model\,C & $\frac{\chi^2}{ \rm dof} = \chi^2_r$ & 1.05 & 0.98 & 0.99 & 1.04 & 0.98  \\
\hline
\textsc{xillverCp}& $\Gamma$ & $1.77\pm0.01$ & $1.76\pm0.03$ &$1.80\pm0.02$ & $1.80\pm0.01$ & $1.82\pm0.02$\\
\textsc{xillverCp}& kT$_{\rm e}$(keV) & $78^{+32}_{-26}$ & $75^{+54}_{-61}$ &$>135$ &$>175$ & $>175$\\
\textsc{xillverCp}&R $_{\mathrm{refl}}$ & $<3.37$ & $1.45\pm 0.62$ & $<0.82$ &$0.16\pm0.04$ & $<0.96$\\
\textsc{xillverCp}&A $_{\mathrm{Fe}}$  & $>3.74$ & $>1.59$ &$1.01^{+0.85}_{-0.25}$ &$0.87^{+0.63}_{-0.16}$ &$<0.50$\\
\textsc{xillverCp}& $\log(N_{\rm H,disk}/\rm cm^{-2})$ &$18.01^{+1.57}_{-0.64}$ &$19.18^{+0.69}_{-1.09}$ & $<16.20$ &$<17.21$& $>15.03$\\
\hline
\hline
\multicolumn{2}{l}{F$_{2-10}(10^{-11}$\,\flux)} & $11.04\pm0.24$ & $7.79 \pm 0.53$ & $9.09 \pm 0.29$ & $8.64\pm0.36$ & $8.49\pm0.27$\\
\hline
\hline
\end{tabular}
\end{table*}
\begin{figure}
    \includegraphics[width=\columnwidth]{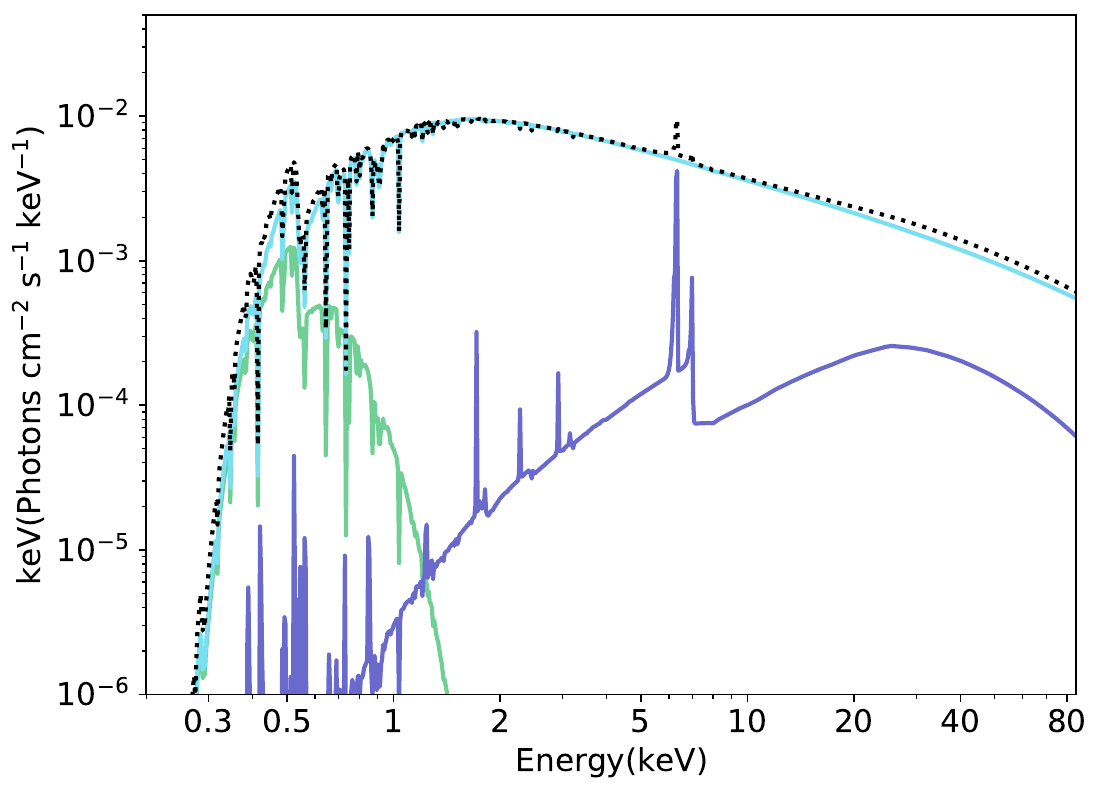}
    \caption{Model\,A (see Sect. \ref{sect:broadband}). The dotted black line represents the total model, which is composed of the primary X-ray emission (light blue line), the reprocessed component (dark blue line), and the soft excess component (green line). Absorption and emission features are included in the model.}
    \label{fig:modelA}
\end{figure}
\begin{figure*}
\centering
\includegraphics[width=0.3\textwidth]{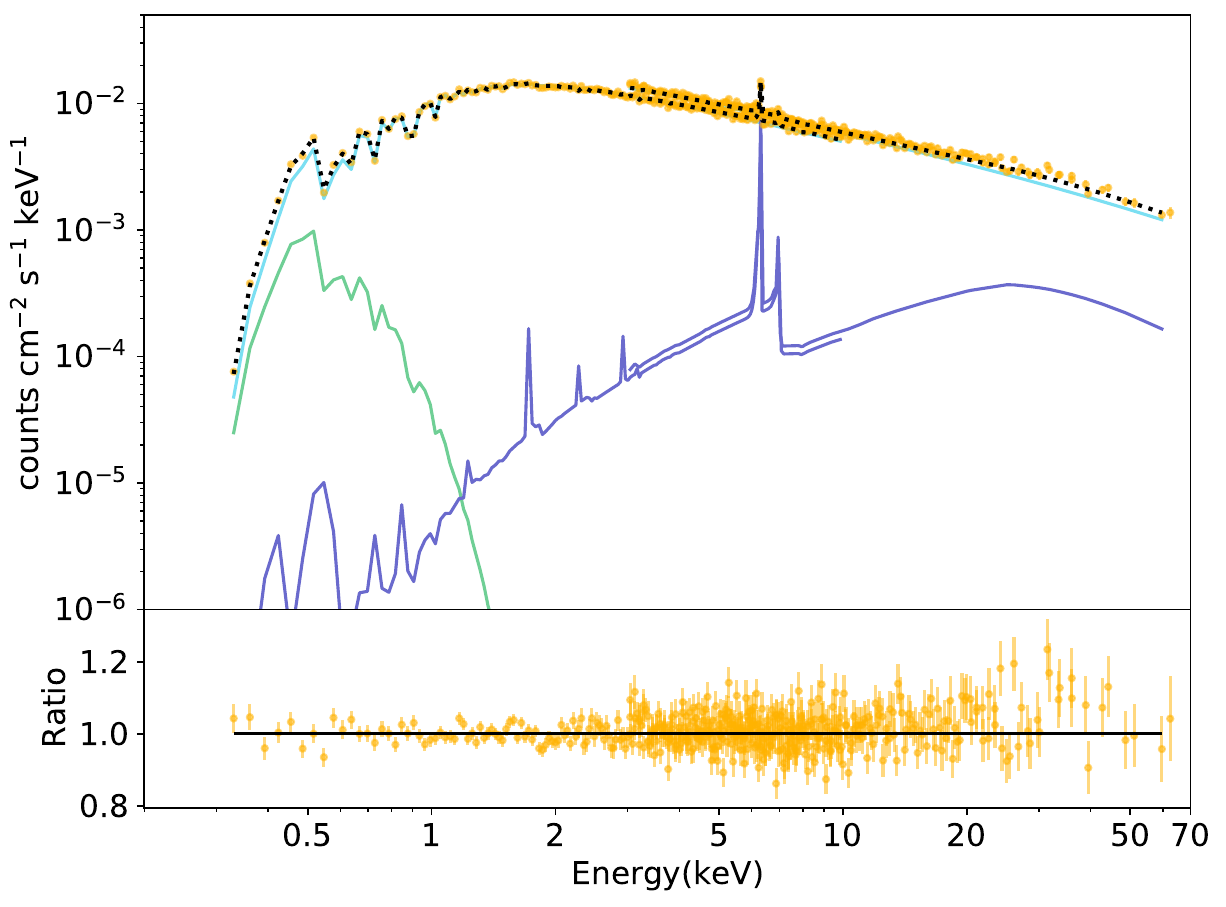}
\includegraphics[width=0.3\textwidth]{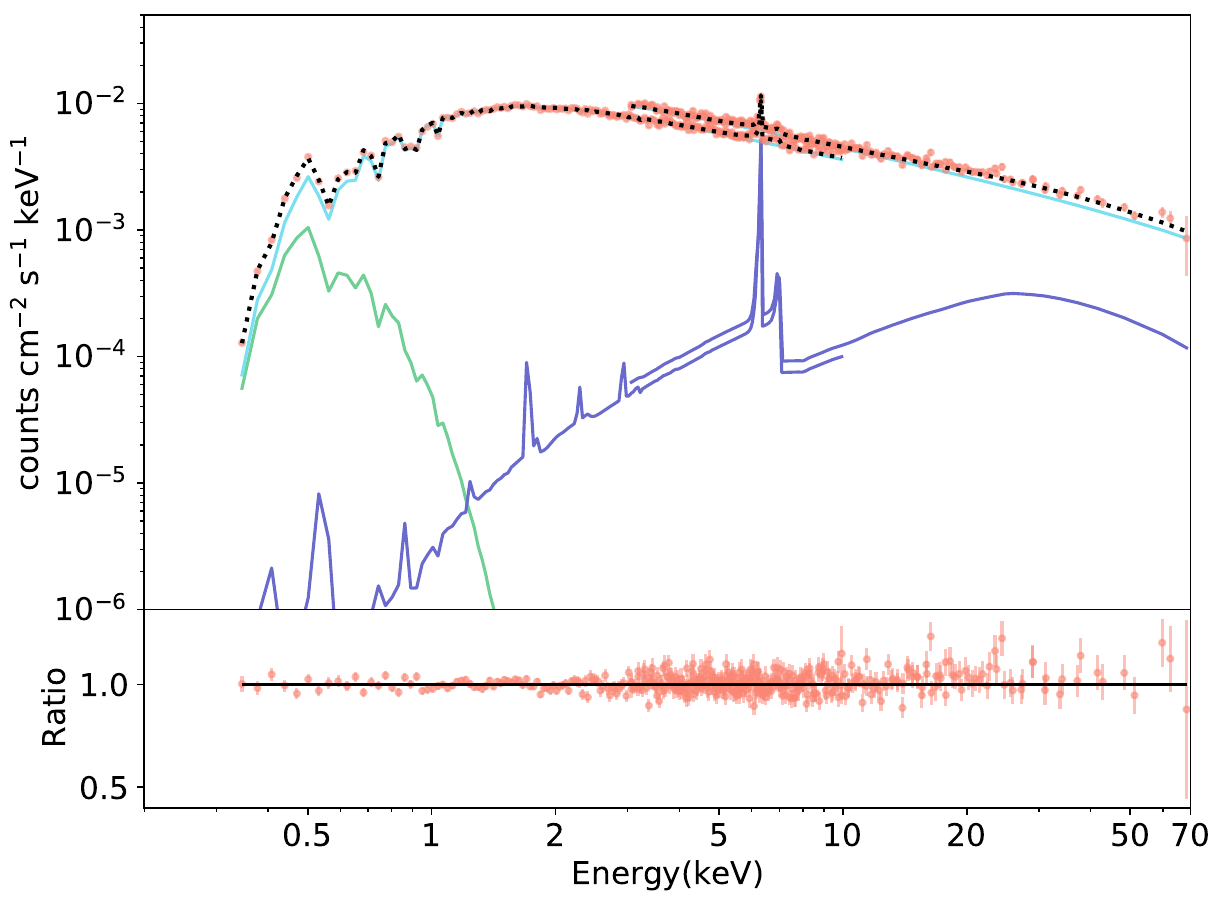}
\includegraphics[width=0.3\textwidth]{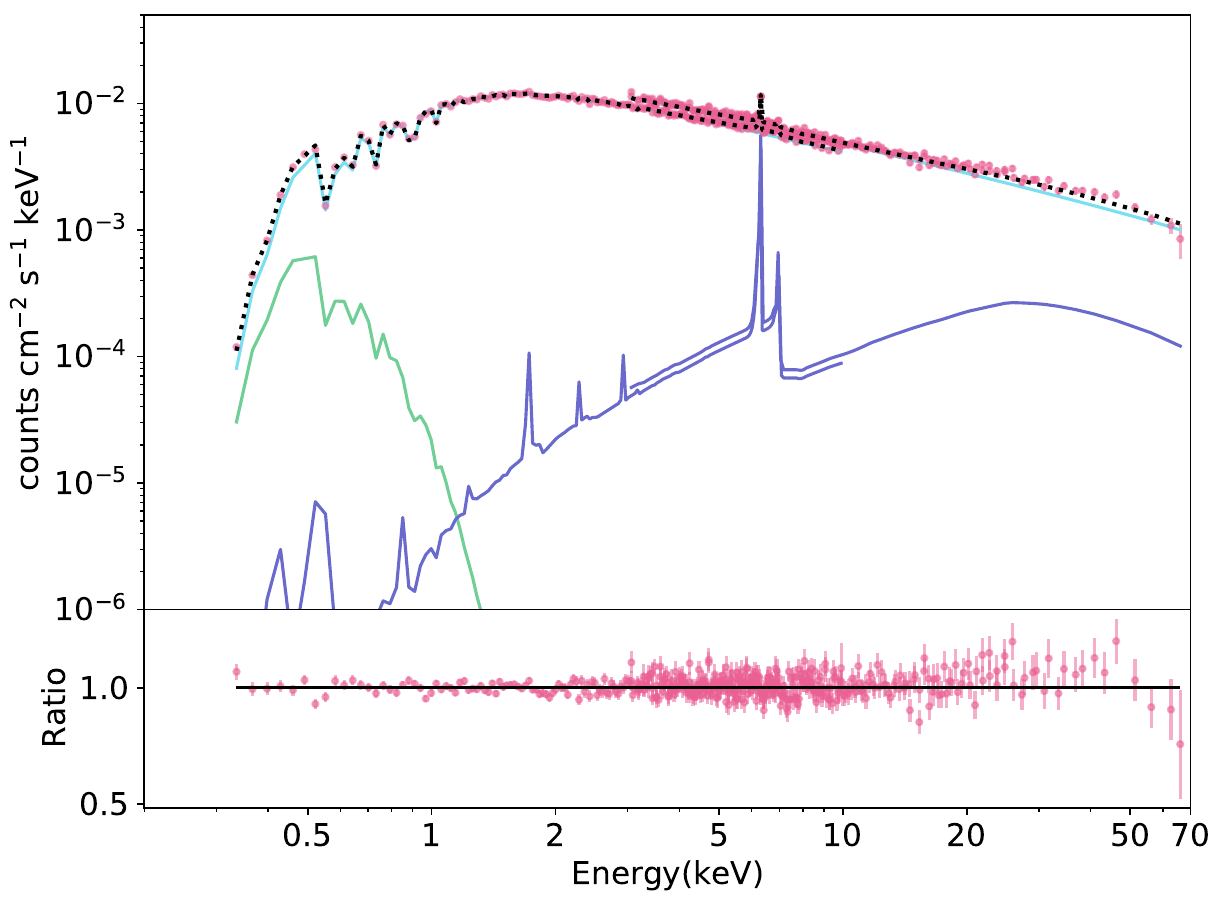}
\includegraphics[width=0.3\textwidth]{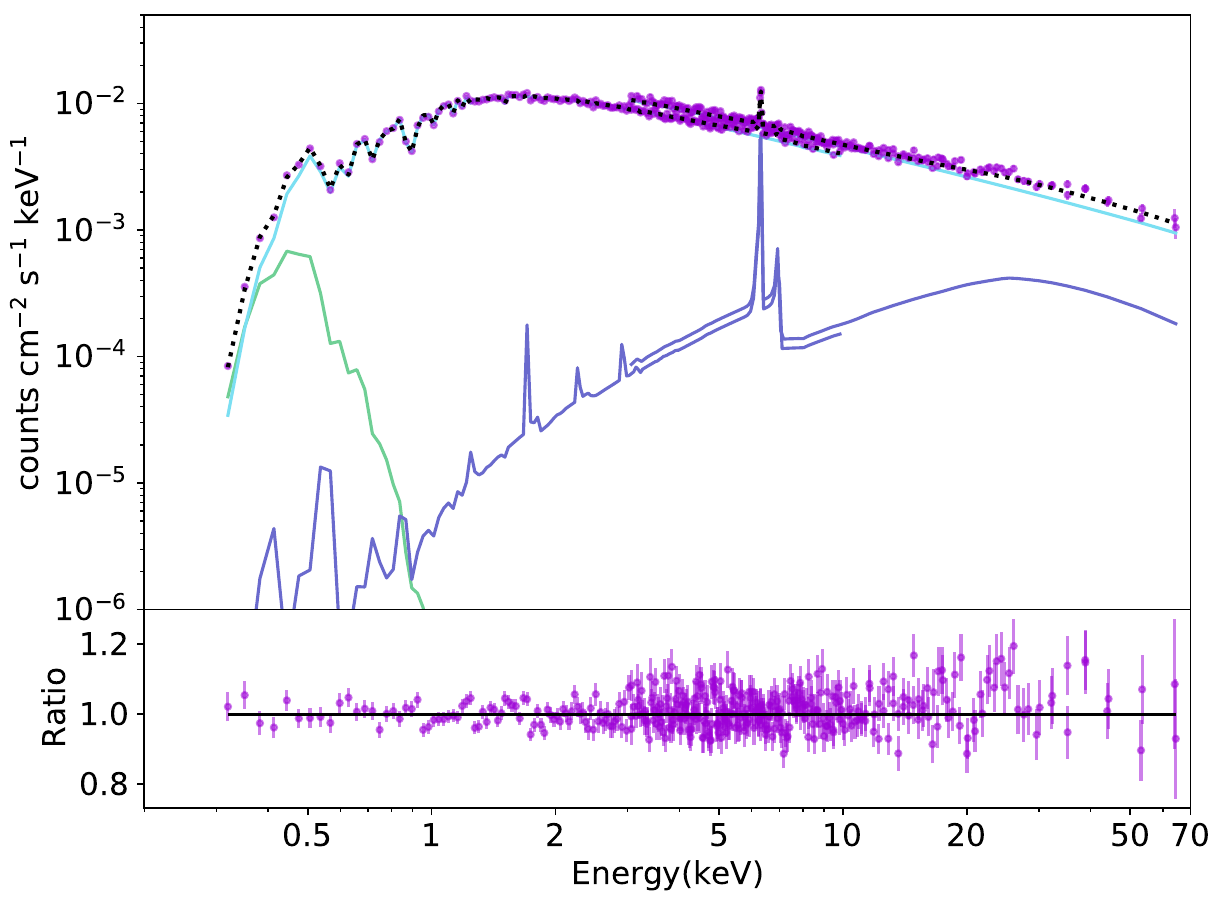}
\includegraphics[width=0.3\textwidth]{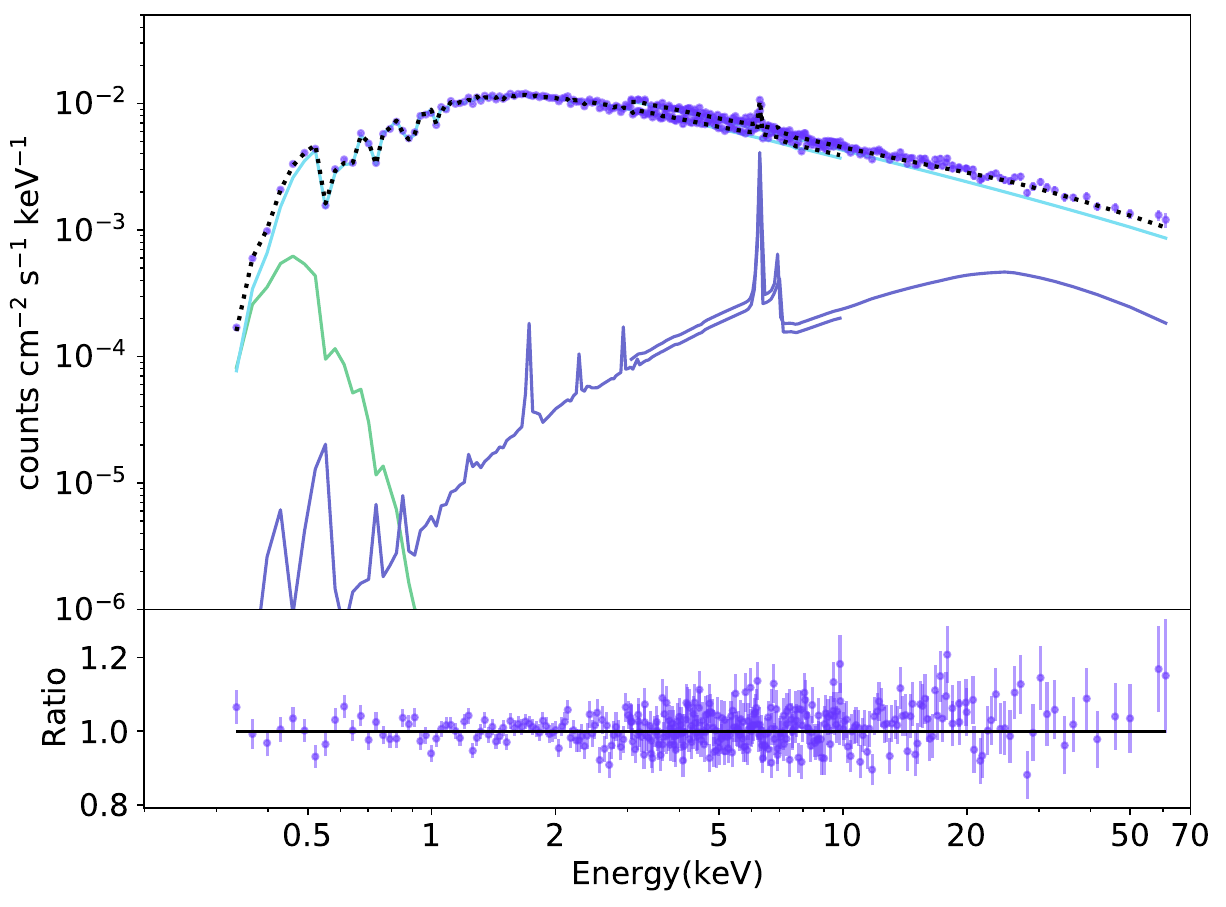}
\caption{Unfolded data plus best-fitting model (upper panels) and ratio residuals (lower panels) for the X-ray broadband observations of IC~4329A when the best-fitting model (Model\,A) is applied. We show all the model components as in Fig. \ref{fig:modelA} (see Sect. \ref{sect:spec} for more details).}
\label{fig:euf_ratios}
\end{figure*}
Finally, we extended the analysis to include the whole 0.3--10\,keV \xmm EPIC-pn spectrum and the 3--75\,keV \nustar FPMA/FPMB spectra for the simultaneous observations (see Table\,\ref{tab:observations_table}). For the spectral fitting, we used the following \textsc{xspec} model: \textsc{tbabs * ztbabs * mtable(WAcomp) * mtable(UFOcomp) * (bbody + cutoffpl + xillver)} (hereafter Model\,A; see Fig. \ref{fig:modelA}), which is the combination of the previous models. In Fig.\,\ref{fig:euf_ratios} we show the unfolded data and ratio residuals for the X-ray broadband observations of IC~4329A when Model\,A was applied.
The data/model ratio of this analysis and the resulting $\chi^2_r$ are shown in the right panels of Fig. \ref{fig:ratios}. We show the data, fitting model, and residuals ratio for all the simultaneous \xmm and \nustar observations in Fig. \ref{fig:cont_spec}. 
Based on the analysis of the \xmm data, we cannot distinguish different components of the Fe \ka emission. We therefore carried out additional tests to determine whether we might understand the origin of the Fe \ka line. The \textsc{xillver} model accounts for the narrow Fe\,\ka emission line, so that we included a Gaussian line for a broad component. This component appeared to be superfluous as its normalization was unconstrained, and we only found an upper limit of $\sim 10^{-7}$\,ph cm$^{-2}$ s$^{-1}$ in all the observations, with an upper limit on the equivalent width in the range of 5--10\,eV. We reported the best-fitting parameters for the X-ray primary continuum ($\Gamma$ and E$_c$) and for the reflection component (A$_{\rm Fe}$ and R$_{\rm refl}$) in Table\,\ref{tab:bb_fit}.

The very low value of the reflection fraction and the faint broad component of the Fe \ka line in the fit of the broadband spectra support the hypothesis in which the disk reflection is weak and the narrow iron \ka signature comes from distant material. To further test this scenario, we replaced the \textsc{xillver} with the \textsc{mytorus} model \citep{2009MNRAS.397.1549M} to reproduce the reflection from distant neutral material (Model\,B1). The inclination angle of the torus was fixed at 30$^{\circ}$. To reproduce the Compton-scattered continuum, we used a table\footnote{\textit{mytorus-scatteredH160-v00.fits}} in which the termination energy was fixed by default in the model and was 160\,keV, the closest value to the cutoff energy found with Model\,A. To reproduce the fluorescent line spectra, we used another table\footnote{\textit{mytl-V000010nEp000H160-v00.fits}} with a line centroid offset $E_{\rm off}=0$\,keV and with the same termination energy as in the previous table. The column densities of the scattered and line components were linked and were free to vary. The normalization of the scattered and line components were tied to the normalization of the primary continuum, that is, the standard \textsc{mytorus} configuration (the so-called coupled reprocessor; see, e.g., \citealt{2012MNRAS.423.3360Y}). The assumed geometry corresponds to a covering fraction of 0.5.

The statistical significance of the fits with Model\,B1 is similar to that of Model\,A. We found a value for the average column density of the torus of $2.6\times 10^{23}$\,cm$^{-2}$, consistent within the errors with the results obtained for all the observations of this campaign (see Table\,\ref{tab:bb_fit}). The photon index values obtained for the primary continuum are consistent within the errors with those obtained with Model\,A, but we were not able to constrain the cutoff energies, most likely because of the degeneracies among the parameters. Moreover, \textsc{mytorus} does not include a cutoff energy, but a termination energy that was fixed to be 160\,keV. We report the best-fitting parameters of Model\,B1 in Table\,\ref{tab:bb_fit}.

Additionally, we tested the reflected emission from a dusty torus with a variable covering fraction (Model\,B2) using the \textsc{RXTorusD}\footnote{\url{https://www.astro.unige.ch/reflex/xspec-models}} model \citep{2017A&A...607A..31P,2023ApJ...945...55R}, the first torus model to consider dusty gas. In these tables, the cutoff energy was fixed to 200\,keV, which is consistent with the value of $E_{\rm c}$ we found in IC~4329A (see Table\,\ref{tab:bb_fit}). The level of the statistical significance observed in the fits using Model\,B2 is comparable to that observed using Model\,B1. The values of the photon index and the equatorial column density of the torus are also comparable within the errors with those of Model\,B (see Table\,\ref{tab:bb_fit}). With Model\,B2, we were able to measure the ratio of the inner to outer radius of the torus, that is, $r/R=0.69\pm 0.02$, and the viewing angle $i\sim 45^{\circ}$. These values were constant in the observations. All the parameters values are reported in Table\,\ref{tab:bb_fit}.

The results obtained with Models\,A, B1, and B2 are consistent with the iron \ka line originating from neutral Compton-thin ($N_H=10^{22}-10^{23}$\,cm$^{-2}$) material that does not produce a prominent Compton reflection.

We searched for possible degeneracies between the fitting parameters in Model\,A, performing a Monte Carlo Markov chain (MCMC) model using the \textsc{xspec-emcee} tool\footnote{\url{https://github.com/jeremysanders/xspec_emcee}}. This is an implementation of the emcee code \citep{ForemanMackey2013} to analyze X-ray spectra within \textsc{xspec}. We used 50 walkers with 10000 iterations each, burning the first 1000. The walkers started at the best-fit values found in \textsc{xspec}, following a Gaussian distribution in each parameter, with the standard deviation set to the delta value of that parameter. In the right panels of Fig. \ref{fig:cont_spec} we show the contour plots resulting from the MCMC analysis of Model\,A applied to the broadband 0.3--75\,keV spectra of IC~4329A.

To directly measure the coronal temperature parameter, we used the \textsc{xillverCp} versions of the \textsc{xillver} tables (Model\,C). This model assumed that the primary emission is due to the Comptonization of thermal disk photons in a hot corona, and the reflection spectrum was calculated by using a more physically motivated primary continuum, implemented with the analytical Comptonization model \textsc{nthcomp} \citep{Zdziarski1996, Zycki1999} instead of a simple cutoff power law. The seed photon temperature was fixed at 50\,eV by default in this model. This value is consistent with the maximum disk temperature expected for a source with $M_{\rm BH}=6.8^{+1.2}_{-1.1}\times10^7 M_{\odot}$ \citep{2023ApJ...944...29B} (i.e., $T_{\rm Wien}=32\pm 10$\,eV). As reported in Table\,\ref{tab:bb_fit}, we were able to constrain the coronal temperature for only two observations. The electron temperature of the corona is thought to be related to the spectral cutoff energy, $E_{c} \sim 2-3 \times kT_{\rm e}$ \citep{Petrucci2000,Petrucci2001}. Assuming this relation, the values we obtained with Model\,C for the electron temperature of the corona are consistent with the cutoff values that we obtained with Model\,A.
\subsection{Reanalysis of previous observations}
\begin{figure*}
    \includegraphics[width=\textwidth]{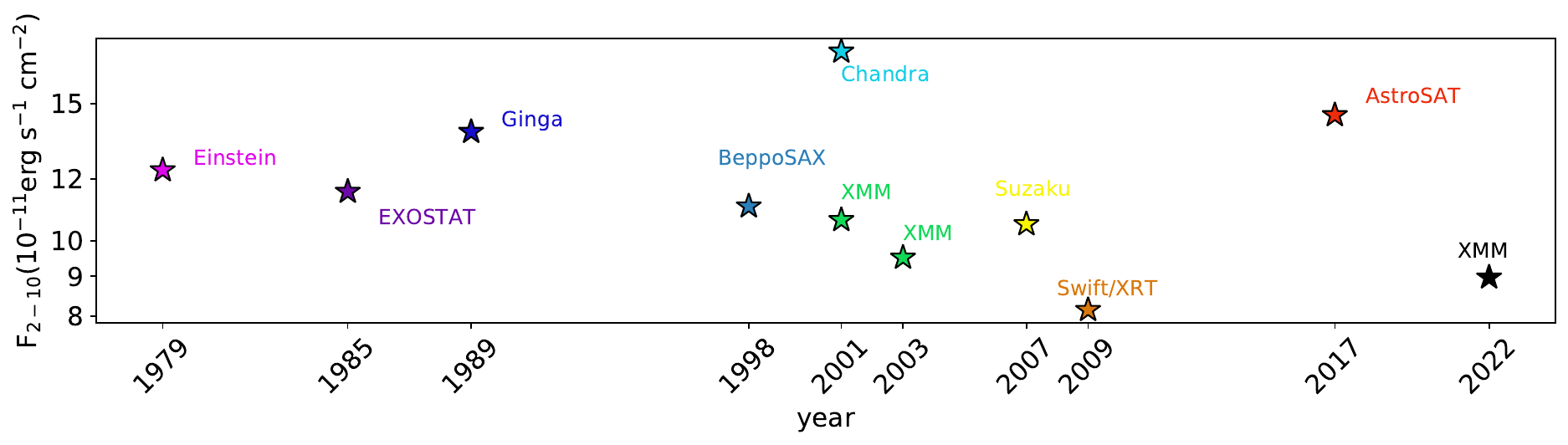}
    \caption{Historical 2--10\,keV light curve of IC~4329A as observed with various X-ray satellites. In chronological order: \textit{Einstein}, 1979 \citep{1989ESASP.296.1105H}; EXOSAT, 1985 \citep{1991ApJ...377..417S}; Ginga, 1989 \citep{1990ApJ...360L..35P}; BeppoSAX, 1998  \citep{2007A&A...461.1209D}; \textit{Chandra}, 2001 \citep{2010ApJS..187..581S}; \xmm 2001, 2003 \citep{2007MNRAS.382..194N}; \suzaku, 2007 \citep{2014MNRAS.442L..95M}; \textit{Swift/XRT}, 2007 \citep{2009ApJ...690.1322W}; AstroSat, 2017 \citep{2021ApJ...915...25T}; \xmm 2022; medium flux of the observations analyzed in this work. }
    \label{fig:2-10_flux}
\end{figure*}

\begin{table}
\caption{Summary of previous \xmm, \nustar, and \suzaku observations of IC~4329A.}
\label{tab:past_observations_table}
\begin{tabular}{llcc}
\hline
\hline
Telescope & Obs. ID & Start Date & Exp\\
 & & yyy-mm-dd & ks\\
\hline
\hline
\xmm & 0101040401 & 2001-01-31 & 13.9\\
\xmm & 0147440101 & 2003-08-06 & 136.1 \\
\suzaku & 0702113010 & 2007-08-01 & 25.4\\
\suzaku & 0702113020 & 2007-08-06 & 30.6\\
\suzaku & 0702113030 & 2007-08-11 & 26.9\\
\suzaku & 0702113040 & 2007-08-16 & 24.2\\
\suzaku & 0702113050 & 2007-08-20 & 24.0\\
\nustar & 60001045002 & 2012-08-12 & 162.4 \\
\suzaku & 0707025010 & 2012-08-13 & 117.4\\
\xmm & 08800760801 & 2018-01-08 & 16.4 \\
\hline
\hline
\end{tabular}
\end{table}
\label{sect:old_data}
\begin{figure}
\centering
\includegraphics[width=0.9\columnwidth]{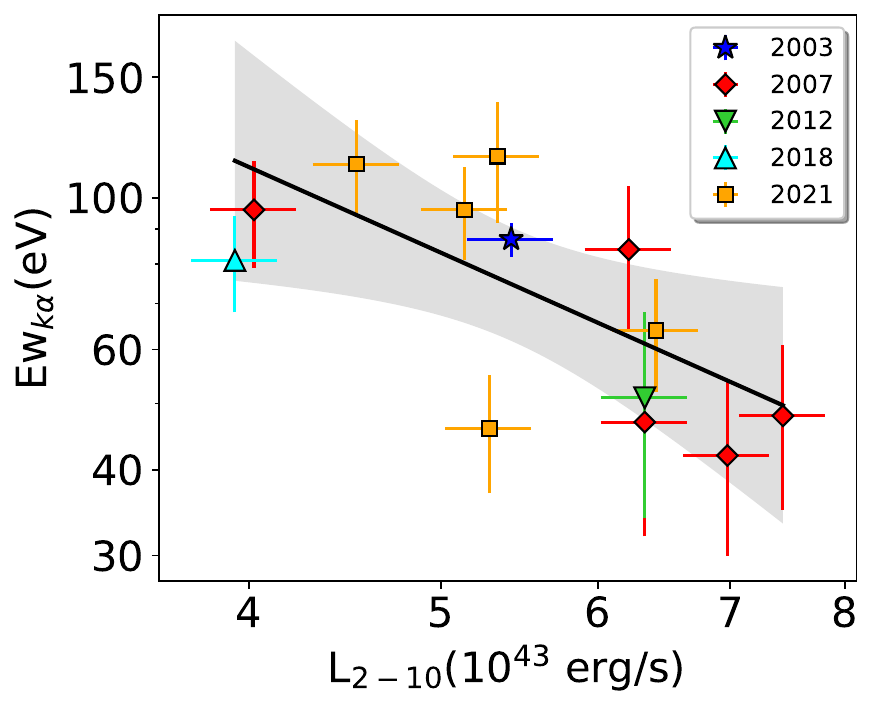}
\caption{Equivalent width (EW$_{K\alpha}$) of the neutral iron line plotted against the 2--10\,keV X-ray luminosity ($L_{\rm 2-10}$). The black line shows the linear regressions, and the shaded black regions represent the combined 3$\sigma$ error on the slope and normalization.}
\label{fig:baldwin}
\end{figure}
\begin{table*}
\caption{Best-fitting parameters of the primary continuum and details of the iron \ka line parameters for the analysis of the spectra of IC~4329A from previous \xmm and \suzaku observations as in Table \ref{tab:past_observations_table}. The errors are at the 90\% confidence levels. The Fe\ka flux and the 2--10\,keV flux are given in units of \flux.}
\label{tab:best_fit_past}
\begin{tabular}{ccccccccc}
\hline
\hline
 Obs.ID & kT$_{\rm BB}$[keV] & $\Gamma$ & $E_{\rm Fe K\alpha}$[keV] & $\sigma_{\rm Fe K\alpha}$[eV] & EW$_{\rm Fe K\alpha}$ [eV]& $\log(F_{K\alpha})$  &$\log(F_{2-10})$ &$\chi^2_r$\\
\hline
\hline
0101040401 & $0.08\pm0.01$   & $1.68\pm0.08$&-              &-           &-            &-               & $-9.46\pm0.55$ &0.97\\
0147440101 & $0.42^*\pm0.01$ & $1.71\pm0.01$&$6.35\pm0.04$  &$96\pm17$   & $87\pm5$    & $-12.12\pm0.03$ & $-10.03\pm0.01$&1.74\\ 
0702113010 & $0.08\pm0.05$   & $1.84\pm0.01$& $6.38\pm0.03$ &$115\pm32$  & $84\pm20$   & $-11.98\pm0.07$& $-9.97\pm0.03$ &1.04\\
0702113020 & $0.03\pm0.02$   & $1.88\pm0.02$&$6.40\pm0.02$  &$82\pm12$   & $48\pm12$   & $-12.32\pm0.07$ & $-9.89\pm0.01$ &1.10\\
0702113030 & $0.07\pm0.01$   & $1.77\pm0.01$& $6.37\pm0.04$ &$71\pm6$    & $42\pm12$   & $-12.08\pm0.08$ & $-9.92\pm0.03$ &1.08\\
0702113040 & $0.07\pm0.01$   & $1.85\pm0.01$&$6.39\pm0.04$  &$78\pm4$    & $47\pm15$   & $-12.15\pm0.10$ & $-9.96\pm0.04$ &1.06\\
0702113050 & $0.06\pm0.01$   & $1.74\pm0.01$&$6.36\pm0.05$  &$67\pm10$   & $96\pm21$   & $-12.41\pm0.12$  & $-10.16\pm0.01$  &1.04\\
0707025010 & $0.06\pm0.02$   & $1.75\pm0.01$& $6.36\pm0.06$ &$77\pm14$   & $51\pm27$   & $-12.04\pm0.03$  & $-9.96\pm0.01$ &1.10\\
08800760801 & $0.11\pm0.05$  & $1.63\pm0.03$& $6.39\pm0.03$ &$<140$      & $81\pm13$   & $-11.91\pm0.03$& $-10.17\pm0.24$&0.93\\
\hline
\hline
\end{tabular}\\
{\raggedright $^*$ The higher value of the kT$_{\rm BB}$ of this Obs.ID is not related to the variability in the soft excess, but it is most likely due to the higher $N_H$ value of the UFO component (see Table\, \ref{tab:x-raywinds}), which could drive the blackbody shape of the soft excess.\par}
\end{table*}
To better understand the origin of the Fe \ka line, the evolution of the spectral parameters of the primary continuum and of the soft energy components (e.g., soft-excess or outflowing components), we compared the results of our campaign with those obtained by previous observations of IC~4329A (see Table \ref{tab:past_observations_table}). This was done by reanalyzing the archival \xmm (2001, 2003, 2018), \nustar (2012), and \suzaku (2007, 2012) data. In Fig. \ref{fig:2-10_flux} we show the 2--10\,keV flux of IC~4329A during the years, including the literature values from observations carried out in the past four decades. During this period, the 2--10\,keV flux of IC~4329A has changed by a factor $\sim 2.1$. The highest state was in 2001, when the flux was F$_{\rm 2-10}=1.75\times 10^{-10}$ \flux, while in 2021, the source reached the lowest state with F$_{\rm 2-10}=8.15 \times 10^{-11}$ \flux. This is lower by $\sim 70$\% than the value of 2001.

We applied the same fitting approach as outlined in Sect. \ref{sect:xmm} to the archival \xmm observations and to the 2007 \suzaku observations, and we used Model\,A (see Sect. \ref{sect:broadband}) to fit the simultaneous \nustar and \suzaku observations of 2012.

The continuum emission of all previous \xmm observations and of the \suzaku observations of 2007 is well fit by a power law and a blackbody component. The latter was included to take the soft excess into account. The simultaneous \nustar and \suzaku observations of 2012 are well fit by Model\,A (see Sect. \ref{sect:broadband}). From the analysis of these simultaneous observations, we found a primary power law characterized by a photon index $\Gamma=1.74\pm0.01$ and an exponential cutoff at E$_c=173\pm13$\,keV, in agreement with what was found by \citet{2014ApJ...788...61B}. We also found an iron abundance A$_{\rm Fe}=1.08\pm0.22$ and a low value of the reflection fraction: R$_{\rm refl}=0.009\pm0.004$. This agrees with the results obtained from our monitoring campaign (see Sect. \ref{sect:broadband} and Table \ref{tab:bb_fit}). We report the values of the best-fitting parameters of the primary continuum for the \xmm and \suzaku observation in Table \ref{tab:best_fit_past}. The data, fitting models, and residuals ratio are shown in Fig. \ref{fig:cont_spec}. Moreover, all these observations, as well as the \suzaku observation from 2012, show a WA and a UFO component, except for the \xmm observation from 2001, for which ionized absorbers could not be confirmed due to the low S/N. The values of the parameters of these components are reported in Table \ref{tab:x-raywinds}. During the 2007 \suzaku monitoring (five observations, each $\sim 5$\,ks long) the spectral parameter values are constant within the errors. We therefore report the average values. For a consistent analysis, we also applied the same model to the {\it XMM-Newton} observation of 2003 as in \citealt{2011ApJ...742...44T} (i.e., a power-law component absorbed by Galactic column density, and an \textsc{xstar} table with a turbulent velocity of 1000\,$\rm km\,s^{-1}$ to the 4--10\,keV band). We found values similar to those reported in \citealt{2011ApJ...742...44T}: $N_H=(1.69\pm0.23)\times10^{22}$\,cm$^{-2}$, $\log(\xi/\rm erg\,\rm s^{-1} \rm cm)=4.23\pm0.82$, $v_{\rm w}/c=0.095\pm0.004$.

Additionally, we found the values of the equivalent width (EW$_{K\alpha}$) of the Fe\ka line obtained from previous observations and from the observations of the 2021 campaign to be inversely correlated with the X-ray luminosity in the 2--10\,keV band. Fitting these two parameters with a linear regression, we found a marginally significant ($2.8\sigma$) anticorrelation with a Pearson correlation coefficient $\rho=-0.664$ (see Fig. \ref{fig:baldwin}). This trend is known as the X-ray Baldwin or Iwasawa-Taniguchi effect (hereafter, IT effect) \citep{1993ApJ...413L..15I}. This inverse correlation is similar to the classical Baldwin effect, in which the equivalent width of the [\textsc{C IV}] $\lambda$1550 emission line is inversely correlated with the UV continuum luminosity \citep{1977ApJ...214..679B}. The IT effect has been found in a large sample of different objects studied with different instruments \citep{2004MNRAS.347..316P,2006ApJ...644..725J,Bianchi2007,Shu2010,Ricci2014}. Several explanations have been proposed for the IT effect: a decrease in the covering factor of the material forming the FeK$\alpha$ line \citep{2004MNRAS.347..316P,2013A&A...553A..29R}, the dependence of the luminosity on the ionization state of the material that produces the line \citep{1997ApJ...488L..91N,2000ApJ...534..718N}, and the variability related to the nonsimultaneous reaction to flux changes in the continuum of the reprocessing material \citep{2006ApJ...644..725J}. One notable distinction we found in the reanalysis of previous observations of IC~4329A is the lack of a distinct relativistically broadened Fe\ka component. Excluding the \xmm observation of 2001 in which it was not possible to detect the Fe\ka line because the S/N ratio of the observation was too low, the other observations show an unresolved symmetric Fe\ka with a line width that is mostly consistent with the \xmm resolution at $6.5$\,keV and a low value of the line width and equivalent width, similar to what was found in the 2021 \xmm\ -\nustar\ monitoring (see Table \ref{tab:best_fit_past}).
\section{Discussion}
\label{sect:discussion}
\subsection{Spectral variability}
\label{sect:gammavsflux}
\begin{figure*}
\centering
\includegraphics[width=0.7\columnwidth]{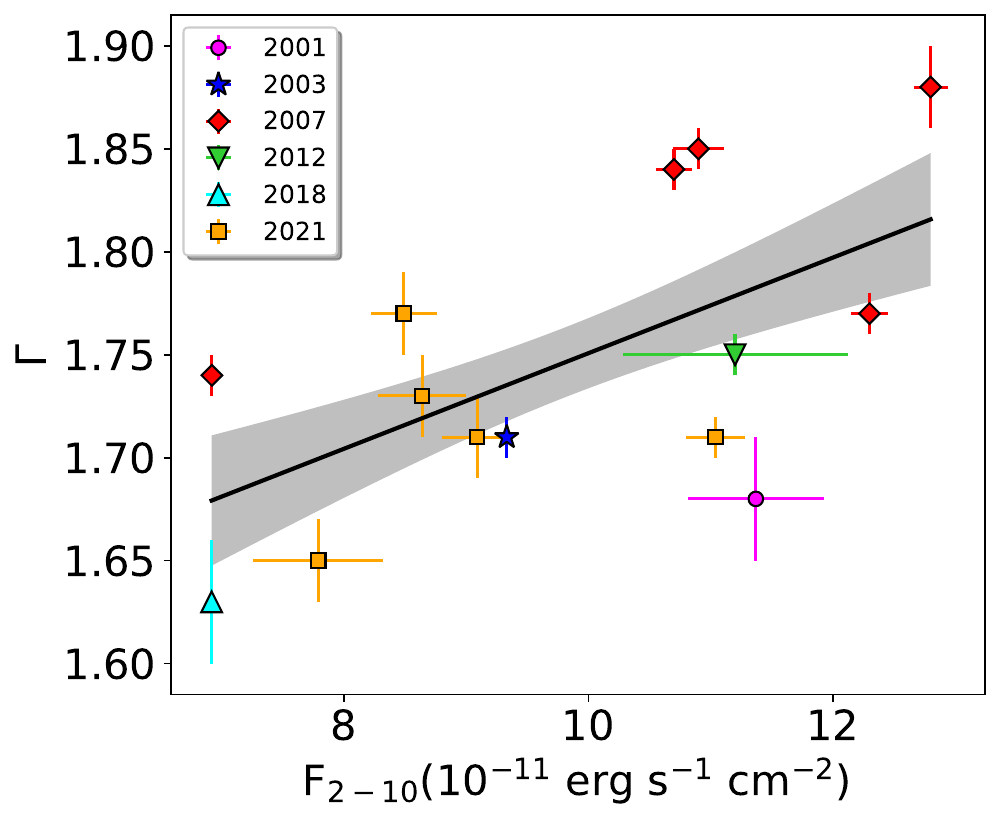}
\includegraphics[width=0.7\columnwidth]{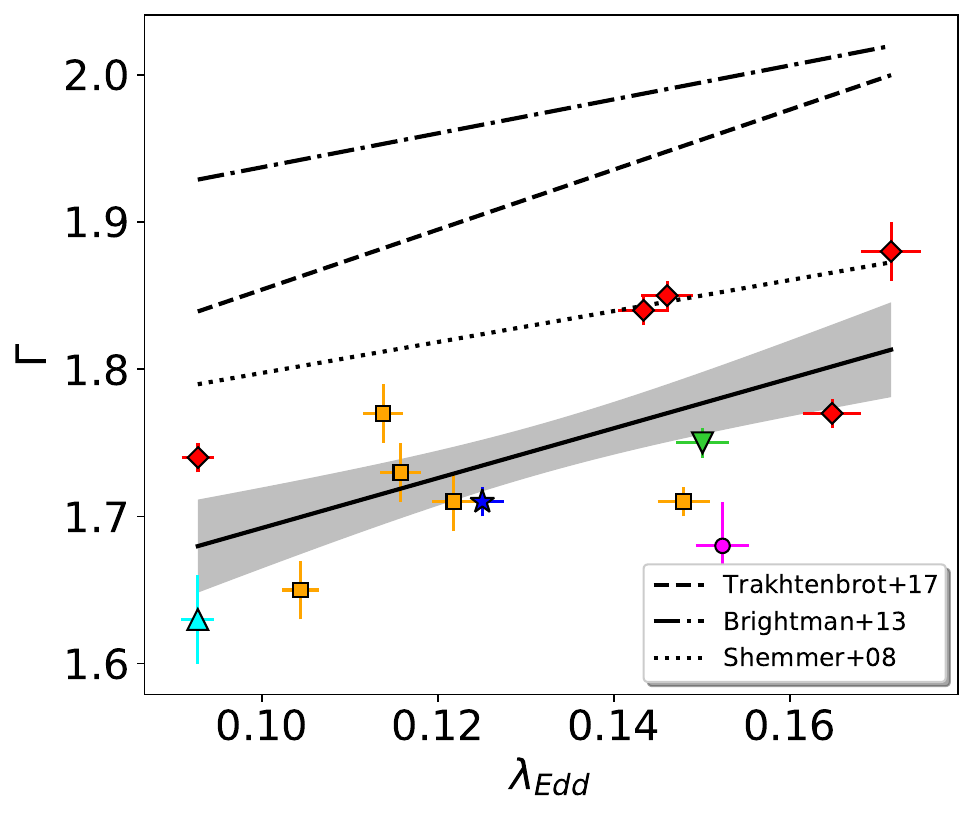}
\includegraphics[width=0.7\columnwidth]{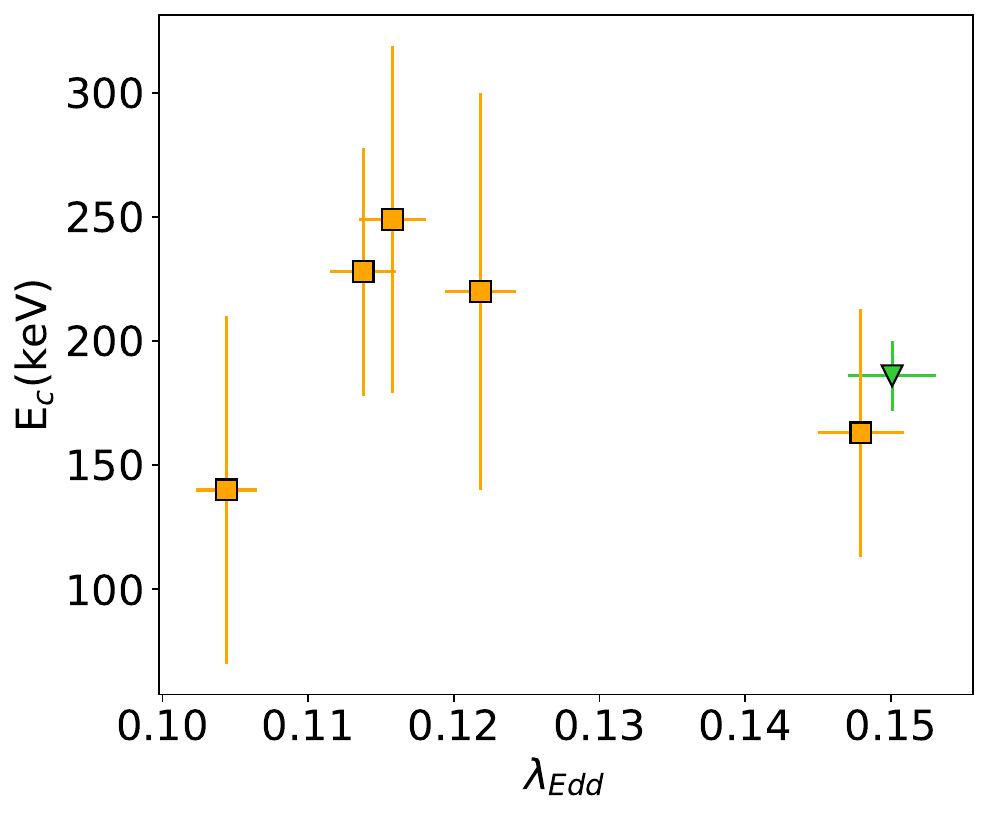}
\includegraphics[width=0.7\columnwidth]{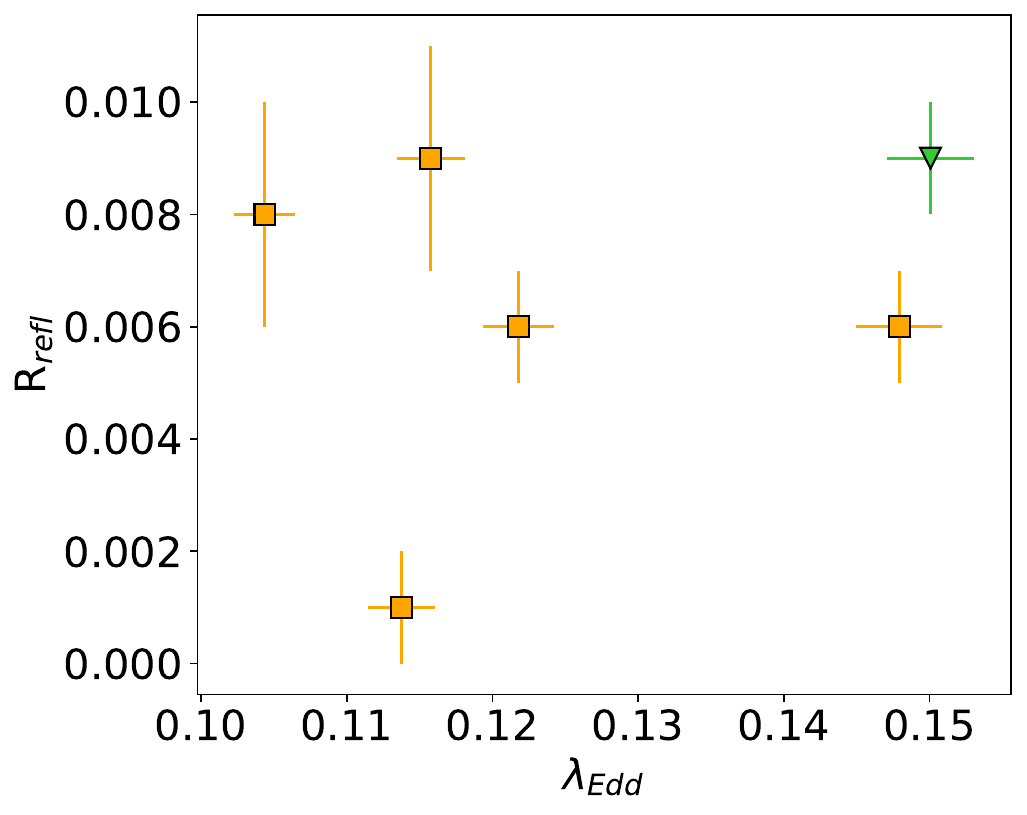}
\caption{Top panels: $\Gamma$ vs. F$_{\rm 2-10\,keV}$ (left) and $\Gamma$ vs. $\lambda_{\rm Edd}$ (right) relation for IC~4329A. The black line shows the linear regression, and the shaded black regions represent the combined 1$\sigma$ error on the slope and normalization. The dotted, dotted dashed, and dashed lines in the right panel represent the  relations from \citet{2008ApJ...682...81S}, \citet{B13} and \citet{2017MNRAS.470..800T}.  Bottom left panel: E$_c$ vs. $\lambda_{\rm Edd}$. Bottom right panel: R$_{\rm refl}$ vs. $\lambda_{\rm Edd}$. The plots include the data from our campaign and the data obtained from archival \xmm, \nustar and \suzaku observations. E$_c$ and R$_{\rm refl}$ cannot be measured when ony \xmm data are available. The Eddington ratio was computed using the bolometric luminosity estimated using the bolometric correction to the 2--10\,keV X-ray luminosity ($\kappa_{2-10}$) from \citet{2010A&A...512A..34L}}
\label{fig:gamma_vs_flux}
\end{figure*}
The continuum 2--10\,keV flux of IC~4329A showed fluctuations over the past 40\,years (Fig. \ref{fig:2-10_flux}), and we also observed significant variability during our campaign. The highest value of the observed 2--10\,keV flux in our monitoring is higher by $\sim$30\%  than the lowest observed value (see Table\, \ref{tab:bb_fit}). Consequently, we aimed to investigate whether the spectral properties of the sources are connected to its flux level. The photon index of the power law shows some evidence for variability between the different observations. Including the values from archival observations, we found that $\Gamma$ shows a moderate correlation with the 2--10\,keV flux with a Pearson correlation coefficient of 0.63, corresponding to a $1-P_{\rm value}$ of 98\% (see the top left panel of Fig. \ref{fig:gamma_vs_flux}). This agrees with the softer-when-brighter behavior that is typically observed in AGN \citep{2009MNRAS.399.1597S,2017MNRAS.470..800T}. Specifically, it was found that as the 2--10\,keV flux increases, the photon index tends to become steeper \citep{2006ApJ...646L..29S}. This could be related to changes in the physical conditions of the accretion disk as the flux increases \citep{1991ApJ...380L..51H}, or to the effect of pair production in the X-ray corona \citep{2018MNRAS.480.1819R}.

Several studies in the past decades have shown that $\Gamma$ is strongly correlated with $\lambda_{\rm Edd}$ \citep{1999ApJ...526L...5L,2004ApJ...607L.107W,2006ApJ...646L..29S,2008ApJ...682...81S,B13,2017MNRAS.470..800T}. We searched for this correlation between the values of the photon index of IC\,4392A found in our monitoring and from archival observations, with the Eddington ratio computed using the bolometric luminosity estimated using the bolometric correction to the 2--10\,keV X-ray luminosity ($\kappa_{2-10}$) from \citet{2010A&A...512A..34L}. We found a moderate  $\Gamma$ vs $\lambda_{\rm Edd}$ correlation (Pearson correlation coefficient $\rho_{\rm corr}=0.62$, corresponding to a $1-P_{\rm value}$ of 97\%; see the top right panel of Fig. \ref{fig:gamma_vs_flux}). The linear regression has a slope value of $1.69 \pm 0.80$, which is consistent within the errors with the values expected from the literature \citep[e.g.][]{2008ApJ...682...81S,B13,2017MNRAS.470..800T}, but with a lower normalization value of $1.52 \pm 0.11$.

This result implies that IC~4329A fits the picture in which the mass accretion rate drives the physical conditions of the AGN corona well, which causes the shape of the X-ray spectrum. Therefore, the cutoff energy is expected to show a correlation with the 2--10\,keV flux/luminosity and with the Eddington ratio. In \citet{2018A&A...614A..37T}, we did not find clear evidence of a significant correlation between $E_{\rm c}$ and $\lambda_{\rm Edd}$ in the sample of bright AGN observed at the time with \nustar. However, the sample was still small and did not allow us to exclude a relation like this. \citet{2018MNRAS.480.1819R}, using the values from the BAT AGN Spectroscopic Survey (BASS) X-ray catalog \citep{Ricci2017}, found a negative correlation in the $E_{\rm c}$ versus $\lambda_{\rm Edd}$ relation, suggesting that the Eddington ratio is the main parameter driving the cutoff energy in AGN. The high-energy cutoff values of IC~4329A are marginally different among the observations, but they are still consistent within the errors with previous literature values \citep{2014ApJ...788...61B} and with the median values of bright nearby AGN \citep{2018MNRAS.480.1819R}. We observe no clear correlation in this monitoring between this parameter and $\lambda_{\rm Edd}$ (see the lower panels of Fig. \ref{fig:gamma_vs_flux}).

In the different observations of our campaign, the reflection fraction is consistent with being constant within the errors (see the lower right panel of Fig. \ref{fig:gamma_vs_flux}). This parameter shows a very low value, consistent with what has been observed in the analysis of \suzaku and \nustar observations of IC~4329A by \citet{2019ApJ...875..115O}. This result, in addition to our analysis of the Fe\,\ka line based on \xmm observations, which shows faint broad features (see Sect. \ref{sect:xmm}), suggests that the reflection from the inner disk is weak. This interpretation is further supported by the fact that the narrow features of the iron line seen in the \xmm observations are consistent with being associated with emission from distant neutral material.

IC~4329A also clearly shows a moderate soft-excess component. However, this component does not vary in the observations. Its temperature ranges from kT$_{\rm BB}=68\pm 5$\,eV to kT$_{\rm BB}=75\pm 6$\,eV, consistent within the error in the observations of this campaign and with the temperature found from analyzing the archival observations (see Table\, \ref{tab:past_observations_table}). The flux of the soft-excess component is constant in the observations of our campaign. It is $(2.31-2.89) \pm 0.75 \times 10^{-4}$ ph\,cm$^{-2}$\,s$^{-1}$.
\subsection{Fe K\texorpdfstring{$\boldsymbol{\alpha}$}{alpha} line}
\label{sect:iron_line}
The central energy of the Gaussian line in our analysis is always consistent within the errors with low-ionization or neutral material in all the observations of this campaign (see Table\ref{tab:kalpha}). IC~4329A is well known to exhibit a strong Fe\,\ka line \citep{1990ApJ...360L..35P} with a narrow core at $6.4$\,keV. A broad component, most likely produced in the inner part of the accretion disk and blurred by general relativistic effects, was found in some X-ray observations \citep{2000ApJ...536..213D,2007A&A...461.1209D,2014ApJ...788...61B,2014MNRAS.442L..95M}. From our modeling, just 50\% of the observations of this campaign seems to require a relativistic broad component of the Fe\,\ka line.

Using the mean value of the measured $\sigma_{\rm Fe K\alpha}$ without considering the upper limits (see Table\,\ref{tab:kalpha}), we computed a full width at half-maximum ($FWHM$) of the iron \ka line: $FWHM_{\rm Fe K\alpha}=9078\pm250$\,$\rm km\,/s$. With this value, we estimated the radius of the iron \ka line emission, assuming that the line width represents the Keplerian velocity emitting material, using the following equation:
\begin{equation}
    R_{\rm Fe K\alpha}=\frac{G\,M_{\rm BH}}{v^2},
\end{equation}
where $G$ is the gravitational constant, and $M_{\rm BH}$ is the mass of the central SMBH \citep{1990agn..conf...57N,2004ApJ...613..682P}, and assuming a spherical geometry and an isotropic velocity distribution. This results in the assumption of $v=\sqrt{3}/2v_{FWHM}$. We obtained $R_{\rm Fe K\alpha}=(4.68\pm0.14)\times10^{-5}$\,pc, which corresponds to $\sim 7 R_{\rm g}$, consistent with the inner radius recovered in 50\% of the observations when fitting the line with the \textsc{diskline} model. However, with the mean value of the $FWHM$ of the iron \ka line obtained by the reanalysis of previous observations, we obtain $R_{\rm Fe K\alpha}^{\rm past}=(0.015\pm 0.005)$\,pc. As in \citet{2006MNRAS.368L..62N,2015ApJ...812..113G}, we can compare the radial position of the iron \ka line emitting region with the optical BLR radius R$_{H\beta}$, inferred from H$\beta$ reverberation studies and with the value of the dust sublimation radius R$_{\rm sub}$ (i.e., the inner radius of the dusty torus; \citealt{2008ApJ...685..147N}). R$_{\rm sub}$ is defined as
\begin{equation}
R_{\rm sub}=0.4\left(\frac{L_{\rm bol}}{10^{45} \rm{erg s^{-1}}}\right)^{0.5}\left(\frac{1500\,K}{T_{\rm sub}}\right)^{2.6}[pc],
\end{equation}
where $L_{\rm bol}$ is the bolometric luminosity, and $T_{\rm sub}$ is the dust sublimation temperature, assumed to be the sublimation temperature of graphite grains ($T\sim 1500$\,K, \citealt{2007A&A...476..713K}). Considering a bolometric luminosity of $\log(L_{\rm bol}/\rm erg\,s^{-1})=45.04$ \citep{2017ApJ...850...74K}, we obtain R$_{\rm sub}=0.42$\,pc while R$_{\rm H\beta}=0.012$\,pc from both \citealt{2009ApJ...697..160B,2023ApJ...944...29B}. This value is consistent with the BLR radius resolved by \citealt{2024arXiv240107676G} ($R_{\rm BLR}=0.013$\,pc). \citealt{2024arXiv240107676G} inferred also an inner BLR radius of $R_{\rm BLR,min}=0.0037$\,pc. Based on these values, the bulk of the Fe \ka emission line of IC~4329A appears to likely originate in the inner accretion disk or in the BLR.
\subsection{X-ray outflows}
\label{sect:outflows}
\begin{table*}
\centering
\caption{Spectral parameters, velocity, upper and lower limits on the radial location, and upper limits on the energetic parameters for each X-ray absorber during the years.}
\label{tab:x-raywinds}
\begin{tabular}{lcccccc}
\hline
\hline
$v_{\rm turb}=100$\,$\rm km\, s^{-1}$ & 2003 & 2007 & 2012 & 2018 & 2021\\
\hline
\hline
$N_{\rm H} (10^{21} \rm cm^{-2})$ & $1.35\pm0.01$ & $1.42\pm0.16$ & $1.03\pm0.37$ & $1.75\pm0.72$ & $1.49\pm0.16$\\
$\log(\xi/\rm erg\,\rm s^{-1} \rm cm)$ &$1.28\pm0.08$ & $1.54\pm0.08$ & $1.78\pm0.03$ & $1.70\pm0.67$ & $1.30\pm0.07$\\
$z_{\rm obs}$&$-0.003\pm0.001$ & $-0.016\pm0.011$&$-0.025\pm0.009$ & $-0.024\pm0.056$ & $-0.011\pm0.005$\\
$v_{\rm w}/c$ & $0.02 \pm 0.01$ & $0.03\pm 0.01$ & $0.04\pm0.01$ & $0.04\pm0.05$ & $0.03\pm0.05$ \\
$r_{\mathrm{min}} [10^{-2} \mathrm{pc}]$ & $1.18 \pm 0.01$ & $0.63 \pm 0.04$ & $0.38 \pm0.01$ & $0.40\pm0.01$ & $0.88 \pm 0.03$\\
$r_{\mathrm{max}} [\mathrm{pc}]$ & $1586\pm13$ &  $828\pm93$ &$657\pm236$ &  $465\pm164$ & $1372\pm147$\\
$\dot{M}_{\rm max}/\dot{M}_{\rm acc}$ & 0.38 & 0.24 & 0.13 & 0.24 & 0.30\\
$\dot{p}_{\rm max}/\dot{p}_{\rm rad}$ &0.80 & 1.27 & 1.21 & 1.38 & 1.56 \\
$\dot{K}_{\rm max}/L_{\rm b,out}$& 0.76 & 2.04 & 2.49 &  2.78 & 2.12\\
P$_{\rm null}$ & 2.15E-6 & 3.25E-7& 4.63E-5&2.74E-5 & 4.39E-6\\
\hline
\hline
$v_{\rm turb}=1000$\,$\rm km\, s^{-1}$ & 2003 & 2007 & 2012 & 2018 & 2021\\
\hline
\hline
$N_{\rm H} (10^{21} \rm cm^{-2})$ & $6.68\pm0.19$ & $2.58\pm0.04$ & $2.29\pm0.22$ & $1.49\pm0.02$ & $1.97\pm0.82$\\
$\log(\xi/\rm erg\,\rm s^{-1} \rm cm)$ &$2.44\pm0.05$ & $2.15\pm0.09$ & $2.15\pm0.03$ & $2.02\pm0.11$ & $2.69\pm0.03$\\
$z_{\rm obs}$&$-0.083\pm0.001$ & $-0.076\pm0.011$&$-0.086\pm0.010$ & $-0.094\pm0.020$ & $-0.177\pm0.016$\\
$v_{\rm w}/c$& $0.102\pm0.001 $ & $0.094 \pm0.011 $ &$0.105 \pm0.011 $ & $ 0.114\pm0.022 $ &$0.207\pm0.019$\\
$r_{\mathrm{min}} [10^{-4} \mathrm{pc}]$ & $6.19 \pm 0.13$ & $7.22 \pm 0.16$ & $5.82 \pm0.12$ & $4.97\pm0.19$ & $1.50 \pm 0.28$\\
$r_{\mathrm{max}} [\mathrm{pc}]$ & $55.27 \pm 3.91$ & $11.94 \pm 1.73$ & $126.12 \pm 12.16$ & $261.48 \pm 3.52$ & $42.28 \pm 17.60$\\
$\dot{M}_{\rm max}/\dot{M}_{\rm acc}$ & 0.77 & 0.44 & 0.30 & 0.20 & 0.39\\
$\dot{p}_{\rm max}/\dot{p}_{\rm rad}$ & 1.63 & 2.72 &  3.38 &  5.35 & 3.78\\
$\dot{K}_{\rm max}/L_{\rm b,out}$& 8.34 & 12.92 & 17.85 &  30.54 & 39.32 \\
P$_{\rm null}$ & 9.56E-8 & 5.34E-9& 2.13E-9&7.26E-8 & 1.42E-8\\
\hline
\hline
\end{tabular}
\end{table*}
During the fitting process, we took ionized absorption into account by including two \textsc{xstar} tables, as outlined in Sect. \ref{sect:xmm}. The resulting parameters for the absorption components obtained by our analysis of the 2021 campaign, together with the results obtained for the archival observations, are listed in Table \ref{tab:x-raywinds}.

Following the same approach outlined in different studies (e.g., \citealp{Tombesi2013,Serafinelli2019,2022MNRAS.509.3599T}), we computed upper and lower bounds on the radial position of the absorbers. The lower limit on the radial location of the outflows can be placed 
by estimating the radius at which the observed velocity equals the escape velocity:
$r_{\mathrm{min}}=2GM_{\rm BH}v^{-2}$. The upper limit is placed using the definition of the ionization parameter $\xi =L_{\rm ion}/nr^2$ \citep{tarter1969}, where $L_{\rm ion}$ is the unabsorbed ionizing luminosity emitted by the source between 1\,Ryd and 1000\,Ryd (1 Ryd = 13.6\,eV), $n$ is the number density of the absorbing material, and $r$ is the distance from the central source. To apply this definition, we need to assume that the thickness of the absorber does not exceed the distance from the SMBH, that is, that the absorbers are somewhat compact \citep{Crenshaw2012,Tombesi2013},
$r_{\mathrm{max}} = L_{\mathrm{ion}}N_{\rm H}^{-1}\xi^{-1},$
with $N_{\rm H}$ and $\xi$ the column density and ionization fraction of the absorber, respectively. We computed the ionizing luminosity using the \textsc{luminosity} task in \textsc{xspec} on the unabsorbed best-fit spectral model (see Sect. \ref{sect:spec} and Sect. \ref{sect:old_data}) between 13.6\,eV and 13.6\,keV for all the observations.

The values of the estimated radial location of the absorber components in IC~4329A are reported in Table\,\ref{tab:x-raywinds}. By comparing these values with the literature, it is possible to see that the outflow component with turbulent velocity 100\,$\rm km\, s^{-1}$ (hereafter Wind\,1) shows the typical upper and lower limits of the distance of the WAs for the type\,1 Seyfert galaxies \citep{Tombesi2013}, while the outflow component with a turbulent velocity 1000\,$\rm km\, s^{-1}$ (hereafter Wind\,2) is within the range of the average locations of UFOs ($\sim 3\times10^{-4}-3\times10^{-2}$ pc; see \citealt{Tombesi2012}).

Regarding the outflow energetics, we computed the mass outflow rate $\dot{M}_{\rm out}=4\pi r N_{\rm H} \mu m_p C_g v_{\rm w}$ \citep{Crenshaw2012}, where $r$ is the radial location of the absorber, $N_{\rm H}$ is the equivalent hydrogen column density, $\mu$ is the mean atomic mass per proton (= 1.4 for solar abundances), $m_p$ is the mass of the proton, $C_{\rm g}$ is the global covering factor ($ \simeq 0.5$, \citealt{Tombesi2010}), and $v_{\rm w}$ is the radial velocity centroid. It is interesting to compare this quantity with the mass accretion rate of the source (see Table\,\ref{tab:x-raywinds}). The typical value of the mass outflow rate for sources accreting below or close to the Eddington limit is $\dot{M}_{\rm out} \gtrsim 5-10\% \dot{M}_{\rm acc}$ both for UFOs and slower outflows \citep{Tombesi2012}. Both X-ray absorbers agree with this scenario, in which the mechanical power is enough to exercise a significant feedback on the surrounding environment.

We also computed the value of the momentum rate of the outflow, that is, the rate at which the outflow transports momentum into the environment of the host galaxy,
$\dot{p}=\dot{M}_{\rm out}v_w$. This quantity is compared in Table\,\ref{tab:x-raywinds} with the momentum of the radiation of the source, which is defined as the ratio of the observed luminosity and the speed of light. For IC~4329A, this is $\log(\dot{p}_{\rm rad}/\rm erg\, \rm cm^{-1})=35.58$. Outflows accelerated through the continuum radiation pressure are expected to have a $\dot{p}_{\rm out}/\dot{p}_{\rm rad} \gtrsim  1$ \citep{King2015}. The median value of this ratio for UFOs is $\sim 0.96$ after the relativistic correction and $\sim 0.64$ without the relativistic corrections \citep{Luminari2020}. The values of $\dot{p}_{\rm out}/\dot{p}_{\rm rad}$ we found for IC~4329A are in the range between $0.80$ and $5.35$. This suggests that radiation pressure accelerates the material to the escape velocity.

The last parameter we computed was the instantaneous kinetic power of the outflow, $\dot{K}=\frac{1}{2}\dot{M}_{\rm out}v_w^2$, and we compared it with the outflowing observed bolometric luminosity of the source. \citet{Tombesi2012} showed that for a significant feedback in the environment of an AGN, a minimum ratio of the mechanical power of the outflow and the bolometric luminosity of $\sim 0.3\%$ for UFOs and $\sim 0.8\%$ for WAs is required. Based on the values found for IC~4329A, the source fits this scenario well, in which the outflowing winds can impress a strong feedback. 
\section{Conclusions}
\label{sect:conclusion}
We have presented the detailed broadband analysis of the simultaneous \xmm - \nustar monitoring of the Seyfert\,1 galaxy IC~4329A that was carried out in 2021. The results of our analysis are listed below.
\begin{itemize}
    \item Our spectral analysis shows that the X-ray broadband spectra of IC~4329A from the 2021 X-ray monitoring campaign are well fit by a model that includes a soft-excess component, a WA, and an UFO, primary emission modeled by a power law with a cutoff at high energy, and a distant neutral reflection component accounting for the Fe fluorescence (see Fig. \ref{fig:modelA} and Sect. \ref{sect:spec}). 
    \item We found no significant variations in the \xmm and \nustar hardness ratios of the source (lower than 10\%; see Fig. \ref{fig:xmm_lc},\ref{fig:nu_lc} and Sect. \ref{sect:timing}) within the single observations ($\sim 20$\,ks). The variability spectra of IC~4329A during the campaign are nearly flat (see Fig. \ref{fig:fvar}). The analysis of the excess variance values of the observations of this monitoring also suggests that IC~4329A shows very weak spectral variations overall (see Fig. \ref{fig:nxs}). 
    \item We found a photon index of the primary power law ranging from $\Gamma=1.65\pm 0.02$ to $\Gamma=1.77 \pm 0.02$ and a cutoff value $E_{\rm c}= 140 - 250$\,keV, which agrees with the results of \citet{2014ApJ...788...61B}. However, with respect to the results from \citet{2014ApJ...788...61B}, we found a lower value for the reflection fraction, R$_{\rm refl}=0.006 - 0.009$, as found by the analysis of \suzaku and \nustar observations of IC~4329A by \citet{2019ApJ...875..115O}. Together with the very faint broad component of the Fe \ka (see Sect. \ref{sect:xmm}), this suggests that the reflection from the inner disk is weak.
    \item Our analysis of the broadband spectra of IC~4329A suggests that if it is present, the narrow core of the iron \ka line likely originates from neutral Compton-thin ($N_H=10^{22}-10^{23}$\,cm$^{-2}$) material that does not produce a prominent Compton reflection. This is consistent with the low reflection fraction found in the analysis. Moreover, in this 2021 \xmm -\nustar\ monitoring, we found an indication of a relativistically broadened iron \ka line in IC~4329A in 50\% of the observations of the monitoring. However, the data from previous IC~4329A observations did not require an asymmetrical Fe \ka component. From an estimate of the position of the iron K emitting region (see Sect.\,\ref{sect:iron_line}), the Fe \ka emission line of IC~4329A appears to likely originate in the inner accretion disk or in the BLR. At the \xmm\ resolution, the different components of the Fe \ka line are unfortunately unresolved. Future observations with a higher-resolution spectrometer (e.g., {\it XRISM}-Resolve) will be pivotal for distinguishing the different components of the low-ionization Fe~K$\alpha$ emission that is indicated in the current data.
    \item Small spectral changes are observed that follow the softer-when-brighter behavior typically observed in unobscured AGN. This is most likely related to changes in the physical conditions of the accretion flow (see Sect. \ref{sect:gammavsflux}). We found a statistically significant $\Gamma$ versus $\lambda_{\rm Edd}$ correlation (see the upper right panel of Fig. \ref{fig:gamma_vs_flux}), indicating that the Eddington ratio may drive the physical conditions of the X-ray emitting corona.
    \item We estimated the distance from the SMBH of the iron \ka line emitting region from the mean $FWHM$ of the line. We find that the Fe \ka emission line of IC~4329A is consistent with originating in the inner disk (see Sect. \ref{sect:iron_line}).
     \item From the analysis of the \xmm and \nustar observations of IC~4329A from this campaign and from archival data, we found X-ray outflows composed of two phases: one UFO, and one WA with a lower velocity (see Sect. \ref{sect:outflows}). From the energetics of these X-ray winds, we conclude that both components are powered by radiation pressure and that they could exert a significant feedback on the AGN environment.
\end{itemize}

\begin{acknowledgements}
      C.R. acknowledges support from the Fondecyt Regular grant 1230345 and ANID BASAL project FB210003. E.S. acknowledges support from ANID BASAL project FB210003 and Gemini ANID ASTRO21-0003. T.K. is grateful for support from RIKEN Special Postdoctoral Researcher Program and is supported by JSPS KAKENHI grant number JP23K13153. This work is based on observations obtained with the ESA science mission \xmm, with instruments and contributions directly funded by ESA Member States and the USA (NASA), the \nustar mission, a project led by the California Institute of Technology, managed by the Jet Propulsion Laboratory and funded by NASA. This research has made use of the \nustar Data Analysis Software (NuSTARDAS) jointly developed by the ASI Space Science Data Center (SSDC, Italy) and the California Institute of Technology (Caltech, USA). We thank the anonymous referee for his/her valuable suggestions.
\end{acknowledgements}

\bibliographystyle{aa} 
\bibliography{bibliography} 

\end{document}